# An anticipated surprise-hazard framework for risky intertemporal choices

Ho Ka Chan[1] and Taro Toyoizumi[1,2]

[1] *Laboratory for Neural Computation and Adaptation, RIKEN Center for Brain Science*
[2] *Department of Mathematical Informatics, Graduate School of Information Science and Technology, The University of Tokyo*

Corresponding authors: hoka.chan@riken.jp (HKC); taro.toyoizumi@riken.jp (TT)

## Acknowledgement

The authors thank Roberta Martino and Viviana Ventre for their helpful discussions on intertemporal decision-making and its potential relation to the anticipated surprise framework. The study was supported by RIKEN Center for Brain Science, RIKEN TRIP initiative (RIKEN Quantum), and JST CREST program JPMJCR23N2.

## Abstract

People frequently deviate from classical utility theories when making risky and intertemporal decisions. While the effects of risk and temporal delay have been extensively studied in isolation, their interplay and underlying theoretical basis remain debated. In this work, we extend our previously proposed anticipated surprise framework to risky intertemporal choices by modeling delayed rewards as outcomes that fail to materialize at a constant hazard rate. This accounts for key empirical findings: diverse patterns of time inconsistency and aversion to timing risk stem from the avoidance of large negative surprises, while conflicting empirical findings on how risk affects temporal discounting emerge from differences in mental representations of outcome resolution. The extended framework offers a new perspective on how various types of risk may interact and provides a flexible foundation for analyzing complex real-life decision-making scenarios.

Keywords: decision-making, intertemporal choice, risk, anticipated surprise, hazard

JEL codes: D81, D91

# 1. Introduction

Decision-making often involves uncertainty in outcomes. For example, outcomes may be delayed or may occur with a specific probability (risk). Human decisions under risk and time delay are often inconsistent with classical economic theories. For instance, time-consistent models assume that an individual's intertemporal preferences remain invariant as time passes (Samuelson, 1937). However, empirical evidence demonstrates that most people exhibit decreasing impatience towards rewards further in the future (Loewenstein & Prelec, 1992; Myerson & Green, 1995). Furthermore, conventional theories long presumed that the effects of risk and delay are entirely decoupled, meaning that the introduction of risk should not alter how a reward is discounted over time (Loewenstein & Prelec, 1992). Nevertheless, recent empirical works have instead shown that risk influences how much people discount delayed outcomes (Blackburn & El-Deredy, 2013; Keren & Roelofsma, 1995; Öncüler, 2000; Stevenson, 1992; Y. Sun & Li, 2010; Ventre et al., 2025).

Several modeling approaches have been developed to account for these anomalous behaviors related to risk and delay. Baucells & Heukamp (2010), Rambaud & Pérez (2020), Vanderveldt et al. (2015) and Yi et al. (2006) proposed utility functions that fit well to experimental data on decisions involving both risk and delay. Other studies suggest that separate psychological processes are involved in decision-making, for example a more instinctive system (System 1) vs a more "rational" system (System 2) (Schneider, 2016), and a certainty-based system vs an uncertainty-based system (Andreoni & Sprenger, 2012). They show that various economic paradoxes can be accounted for by considering the interaction of these systems.

These works consider risk and delay as separate entities. However, empirical research indicates that people exhibit similar behavioral patterns when facing them (Green & Myerson, 2004; Takahashi et al., 2012; Zhou et al., 2019). For instance, individual preferences may shift both when the delay to an outcome is lengthened by a uniform amount, and when the probability of receiving an outcome is reduced by a constant proportion across all available options (Green & Myerson, 2004; Zhou et al., 2019), with both phenomena obeying the Weber-Fechner law for perception (Takahashi et al., 2012). These commonalities have inspired theoretical frameworks that unify risk and delay. Myerson & Green (1995) and Rachlin et al. (1991) suggested that when evaluating risk, people mentally simulate repeated attempts at obtaining a targeted outcome, each separated by a time buffer. Because early simulated attempts may instead generate other outcomes (due to risk), the targeted outcome is perceived to only arrive after a delay. On the other hand, Halevy (2008), Myerson & Green (1995) and Sozou (1998) hypothesized that when people are promised a delayed reward, they implicitly assume that the reward may fail to materialize because of unforeseen events (in this work, these events are referred to as 'hazard', and models that assume the existence of hazard, 'hazard models'), thereby treating delay as risk.

While these conceptualizations provide a vital foundation for relating risk and delay, the direct interaction between them, such as how risk modulates temporal discounting, remains largely unexplored. The mechanisms behind the evaluation of choices with complex structures, such as compound lotteries and timing risk, are also not addressed.

Neuroimaging studies (Luhmann et al., 2008) suggest that people mentally simulate plausible future events associated with delayed outcomes. The anticipated surprise (AS) framework proposed in our previous work (Chan & Toyoizumi, 2024) assumes that such mental simulation underpins option evaluation. In the AS framework, the anticipation of a future event unfolding updates the future state value, generating a reward prediction error ('surprise'). The total surprise accumulated after simulating the entire event space, which encompasses all sampled paths of outcome resolution, contributes to the utility of the option. This raises a natural question: what events might a decision maker (DM) encounter during outcome resolution that influence the evaluation of a delayed reward? Hazard models provide a plausible answer: hazard—unforeseen events that lead to the loss of reward. By constructing outcome resolution paths that explicitly represent the occurrence or non-occurrence of such hazards, the resulting ASH model can account for decision-making involving not just risk (Chan & Toyoizumi, 2024) but also delay, as well as their combination. We hypothesize that this model can provide a new perspective on various empirically observed behaviors across these domains. To avoid confusion, we will refer to hazard-induced risk as 'hazard risk' and explicitly stated (probabilistic) risk as 'probabilistic risk'.

In this work, we first propose a form of utility function based on anticipated surprise that qualitatively reproduces empirical preferences in decision-making under probabilistic risk (Section 2). Next, we describe how reward delay can be modeled by incorporating hazard into the AS framework (Section 3). We then study the properties of the resulting AS-hazard (ASH) model for decision-making involving different types of risks. In Section 4.1, we show that the ASH model can reproduce rich patterns of time inconsistency observed empirically. In Section 4.2, we illustrate how the model addresses uncertainty in delay (timing risk), and show that the ASH model accurately predicts a general aversion to timing risk. In Section 4.3, we examine decision-making scenarios in which both probabilistic risk and delay are involved, proposing that DMs may construct different mental representations of the outcome resolution process as a result of cognitive constraints and biases. We demonstrate that these distinct representations give rise to opposite effects of probabilistic risk on temporal discounting in the ASH model, offering a plausible explanation for the seemingly contradictory empirical findings on this topic. Finally, we compare our framework with previous models (Section 5.1), discuss the limitations of the ASH model and possible workarounds (Section 5.2), propose plausible links to neurophysiological findings (Section 5.3), and suggest potential extensions and applications of the model (Section 5.4).

## 2. The AS framework and a proposed utility function

The AS framework was first introduced in our previous work (Chan & Toyoizumi, 2024). Here, we briefly summarize the core formulation of the model. We define the expectation error $z$ for an outcome $x$ in an option as the difference between $x$ and the expected value across all outcomes within that option, i.e. $z = x - \mathrm{E}(x)$, where $\mathrm{E}(.)$ denotes the expected value.

For options whose outcomes are resolved in a single stage, the surprise value $\Delta$ is given by

$$\Delta = \mathrm{E}(\delta(z)), \tag{1}$$

$$\text{where } \delta(z) = \begin{cases} f(z) \text{ for } z \geq 0 \\ -kf(|z|) \text{ for } z < 0 \end{cases}. \tag{2}$$

In eq. (2), $f(z)$ is an increasing convex function defined for $z \geq 0$, and $k > 1$ is the loss aversion factor. To simplify analysis throughout the work, we adopt the power-law form $f(z) = z^{\alpha}$ with $\alpha > 1$, though the choice of $f$ does not alter our results qualitatively.

For an $n$-stage option, outcome resolution is modeled as a stochastic process, where $s(i)$ denotes the state at stage $i$. We define $S(i) = (s(1), s(2), \dots, s(i))$ as the trajectory of the state path up to stage $i$. Accordingly, the expectation error at stage $i$ is given by:

$$z(i) = \mathrm{E}(x|S(i)) - \mathrm{E}(x|S(i-1)), \tag{3}$$

where $\mathrm{E}(.|S(i-1))$ denotes the expected value given $S(i-1)$, evaluated by averaging over all possible trajectories $\{S(i)\}$ according to the trajectory probability $p(S(i)) = \prod_{i'=1}^{i} p(S(i')|S(i'-1))$. Here, $z(i)$ represents the change in expected outcome as the simulated state path is updated, analogous to a temporal-difference prediction error.

The surprise value $\Delta_i$ for stage $i$ is obtained by averaging the surprise over all trajectories $S(i)$

$$\Delta_i = \mathrm{E}(\delta(z(i))). \tag{4}$$

The total surprise for the option is computed by summing the surprise generated across all stages

$$\Delta = \sum_{i=1}^{n} K_i \Delta_i. \tag{5}$$

As in our previous work, we assumed $K_i = 1$ for all $i$, except where stated otherwise.

In our previous work, we only compared options with nearly identical expected values. In such situations, we proposed that choice is based entirely on $\Delta$, since any effect of expected value on utility is equivalent across options. In this work, however, we compare options with different expected values, and therefore require a utility function that accounts for the contribution of both the expected value and $\Delta$. More specifically, we define the utility function $U$ as

$$U = U_0 g(\Delta), \tag{6}$$

where $U_0 = \mathrm{E}(x)$ is the expected value across the outcomes in an option, and $g(\Delta)$ is continuous and increasing in $\Delta$, with $g(\Delta) \geq 1$ for $\Delta \geq 0$ and $0 < g(\Delta) < 1$ for $\Delta < 0$. Under this definition, the surprise $\Delta$ invokes a correction to the expected value, or more broadly speaking, the 'expected utility' in the classical sense. A positive surprise ($\Delta \geq 0$) boosts the utility, and a negative surprise ($\Delta < 0$) discounts it. We further assume that $g$ is smooth except at $\Delta = 0$. The option with the highest $U$ is chosen.

We also consider the following specific form of $g(\Delta)$ in later analyses:

$$g(\Delta) = \begin{cases} k_1 \Delta & \text{for } \Delta \geq 0 \\ \frac{1}{1+k_2|\Delta|} & \text{for } \Delta < 0 \end{cases}. \tag{7}$$

The form of $g$ in eq. (7) enables the model to qualitatively reproduce the classical S-shaped utility function observed empirically and generated by prominent descriptive models (Kahneman & Tversky, 1979) for a single-stage binary gamble for gain, $(1, p; 0, 1-p)$ (outcome 1 with probability $p$ and 0 with probability $1-p$), as shown in Figure 1. Another motivation for this form of $g$ is that it produces rich time inconsistency patterns consistent with empirical observations, which will be discussed in Section 4.1.

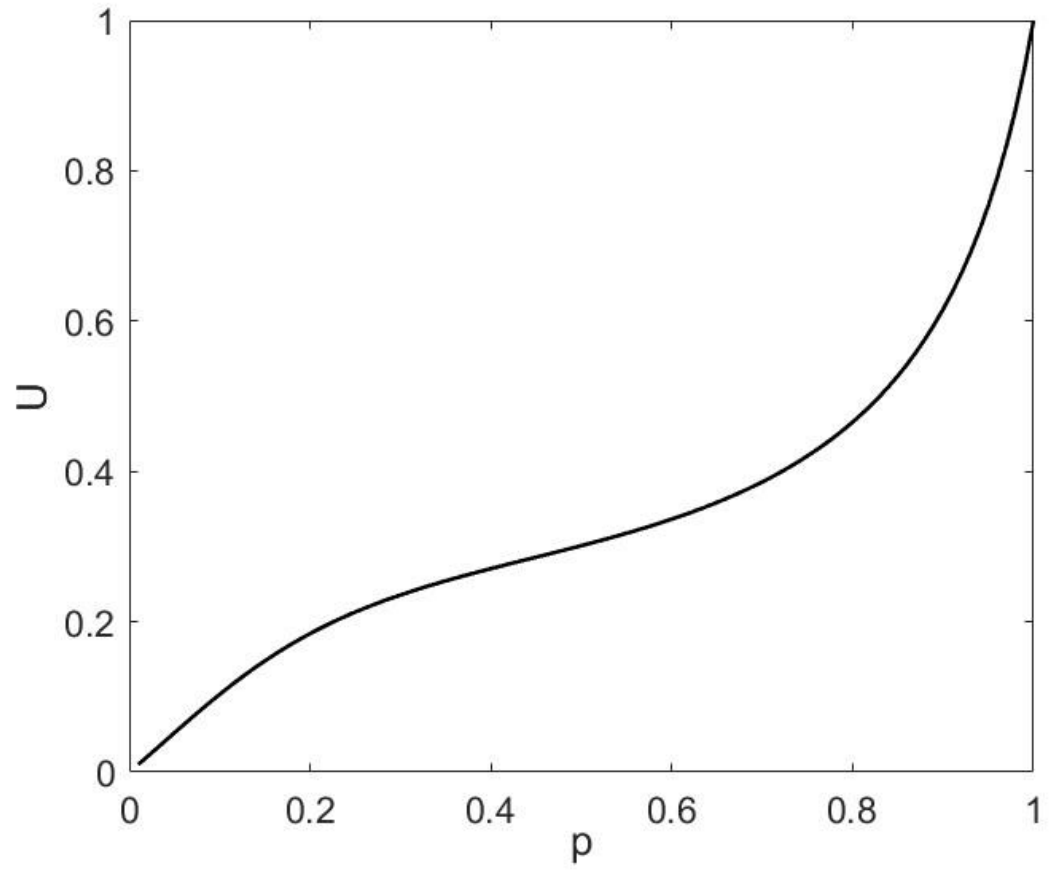

Figure 1: The utility $U$ vs $p$ in a binary gamble. The loss aversion factor $k = 3$, $\alpha = 1.6$ and $k_1 = k_2 = 2$.

In this work, we do not study loss temporal discounting or magnitude effects. Accordingly, the options we analyze are restricted to the gain domain with normalized outcomes, in which eq. (6) is well behaved. It should be noted, however, that directly applying these equations in other domains may lead to erratic predictions due to sign and scaling issues. Please refer to Section 5.2 and Appendix A4 for further explanations of these issues and a simple workaround, and Section 5.5 for potential extension of the model to the loss domain. Furthermore, the analyses in later sections involve only negative surprise, so the form of $g$ for $\Delta \geq 0$ in eq. (7) is irrelevant for subsequent findings and is included only for completeness.

# 3. Representation of reward delay as hazard within the AS framework

In intertemporal decision-making, a DM evaluates options that promise a positive outcome (reward) after a delay. In hazard models (Halevy, 2008; Sozou, 1998), this is represented by 'hazard risk': the possibility that the promised future reward may fail to materialize and become permanently lost. Under the assumption of a constant hazard rate, the DM's mental representation of an option containing a delayed reward can be illustrated by the branching scheme in Figure 2.

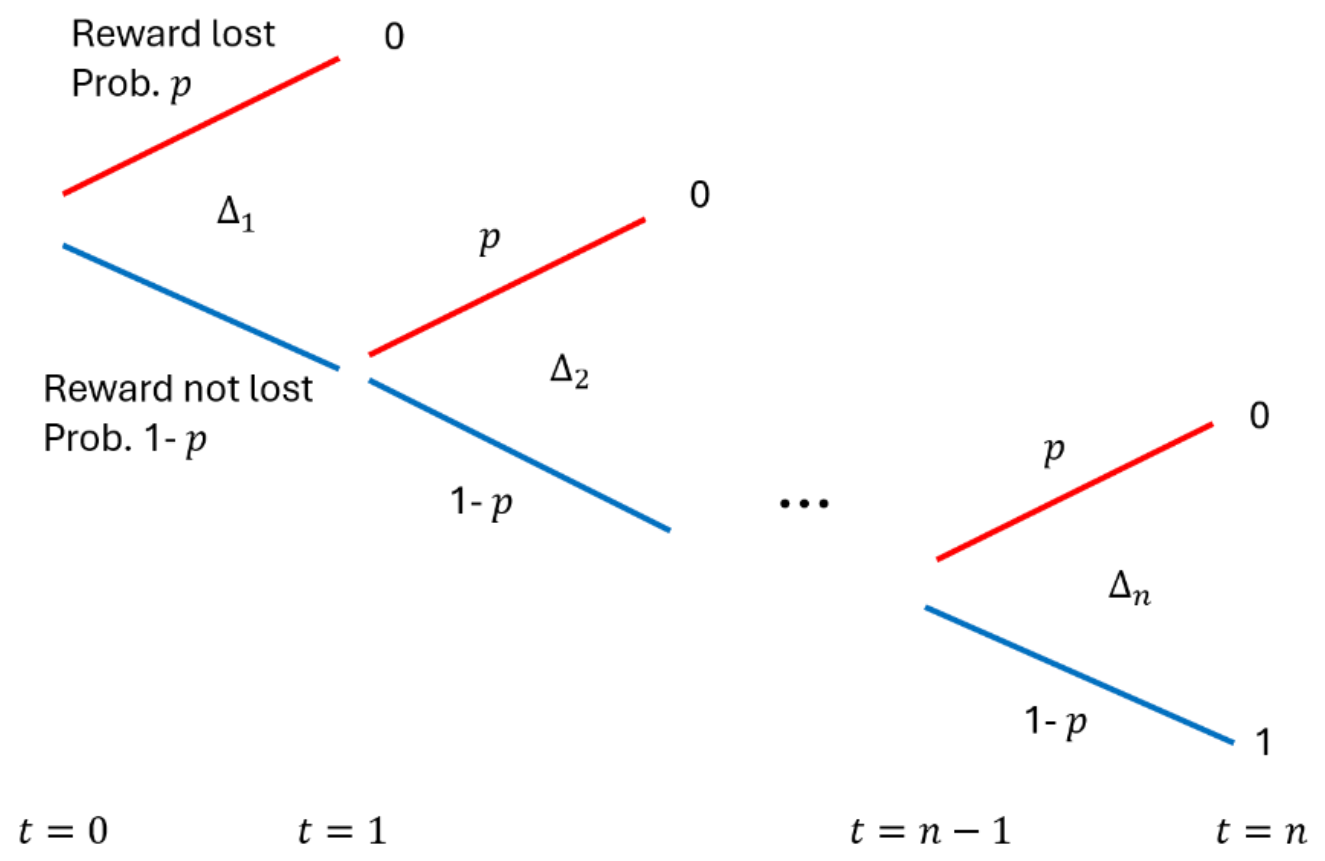

Figure 2: The branching scheme for an option with a delayed reward, in which the reward may be lost due to hazard.

In Figure 2, the DM is promised a reward of size 1 after $n$ time steps. At each time step, there is a probability $p$ that it is lost due to hazard (red branches), yielding an outcome of 0. If it survives all $n$ steps (blue branches), the DM receives the reward, yielding an outcome of 1. As explained in Section 2, rewards are always normalized to 1 throughout this work.

By eq. (4), the surprise generated at each time step $t$ is given by

$$\Delta_t = (1-p)^{t-1}(-C(1-p)^{\alpha(n-t+1)}), \tag{8}$$

where $C = kp - p^{\alpha}(1-p)^{1-\alpha}$, independent of $t$. Since $k > 1$ and $\alpha > 1$, $C$ is positive unless $p$ is large. The term $(1-p)^{t-1}$ corresponds to the probability of reaching time step $t$ without encountering hazard, while the term $-C(1-p)^{\alpha(n-t+1)}$ represents the combined surprise generated at time step $t$ from both the positive outcome (proceeding to the next time step, or receiving the reward at the final step) and the negative outcome (encountering hazard). Unsurprisingly, the negative surprise dominates, since each pair of branches at a given time step corresponds to a probabilistic binary gamble with a small probability of a negative outcome, and aversion to such gambles is widely observed empirically (Kahneman & Tversky, 1979) and reproduced under the AS framework.

The total anticipated surprise of the option in figure 2 ($\Delta_{\text{haz}}$) is given by summing the surprise in eq. (8) across $t$ (See eq. (5)). By reindexing the summation ($t \rightarrow n - t + 1$), we obtain:

$$\begin{aligned}\Delta_{\text{haz}} &= -C(1-p)^{\alpha n} \textstyle\sum_{t=1}^{n} (1-p)^{(\alpha-1)(1-t)} \\ &= -C(1-p)^{n} \textstyle\sum_{t=1}^{n} (1-p)^{(\alpha-1)t} \\ &= -Cq^{n+\alpha'} \frac{1-q^{n\alpha'}}{1-q^{\alpha'}}\,,\end{aligned} \tag{9}$$

where $q = 1 - p$ and $\alpha' = \alpha - 1$.

The expected value of the option is $q^n$, and its utility can be computed using eq. (6) and (7). Because the reward is normalized to 1, this utility is equivalent to the discount factor, i.e. the proportion of the reward's value after accounting for hazard risk relative to its nominal value. From this point onward, $U$ and the discount factor will be used interchangeably.

Although derived under a discrete-time formulation, $\Delta(n)$ in eq. (9) is smooth and well-defined for all continuous delays $n \geq 0$. We therefore adopt this continuous extension hereafter, allowing for non-integer delay $n$ in subsequent analyses.

As a final note, one may ask how the hazard probability $p$ at each time step and the length of a time step in Figure 2 are determined. Intuitively, they depend on factors such as the DM's perception of the environment and their internal state. A precise characterization is beyond the scope of this work and represents an interesting direction for future research. Please refer to Section 5.2.2 for a brief discussion. Most importantly, the results presented in the next section do not rely on any specific choice of $p$ and time-step length.

# 4. Properties of the ASH model

## 4.1 Rich patterns of time inconsistency for basic intertemporal decisions

People empirically exhibit time inconsistency, or more specifically, decreasing impatience for intertemporal decisions described in Section 3 (Loewenstein & Prelec, 1992; Myerson & Green, 1995), which means they prefer an option delivering a reward with a delay $n$, over another option delivering a larger reward with a longer delay $n'$, but reverse their preference if an additional delay, $n''$, is added to both options (Koopmans, 1960). This can be written in terms of the discount factor $U$ as

$$\frac{U(n+n')}{U(n)} < \frac{U(n+n'+n'')}{U(n+n'')}, \tag{10}$$

for $n, n', n'' > 0$.

To examine whether the ASH model also predicts decreasing impatience, it is useful to consider the model's mechanisms for determining $U$. The first mechanism is the reduction of the expected value due to hazard, which by itself produces exponential discounting and therefore does not generate time inconsistency (Myerson & Green, 1995; Sozou, 1998). The second and the third mechanisms are the accumulation of (negative) surprise incurred by hazard and the transformation of the surprise by $g$. Although it might seem obvious that the hyperbolic form of $g$ is sufficient to cause decreasing impatience, we will show below that $\Delta$ plays an important role in enabling the model to reproduce more nuanced patterns of time inconsistency. From eq. (10), decreasing impatience requires $\frac{U(n+n')}{U(n)}$ to be strictly increasing with $n$ for all $n$ and $n' > 0$. Taking the logarithmic derivative yields $\frac{d}{dn}\ln\left(\frac{U(n+n')}{U(n)}\right) = \frac{U'(n+n')}{U(n+n')} - \frac{U'(n)}{U(n)}$. Decreasing impatience is thus equivalent to $\frac{U'(n)}{U(n)}$ being strictly increasing in $n$.

The discount factor is given by $U(n) = U_0(n)g\big(\Delta(n)\big)$, where $U_0(n) = q^n$ represents the baseline reduction in expected value due to hazard. Computing $\frac{U'(n)}{U(n)}$ gives

$$\frac{U'(n)}{U(n)} = \ln q + h(\Delta)\frac{d|\Delta|}{dn}, \tag{11}$$

where $h(\Delta) = \frac{1}{g(\Delta)}\frac{dg(\Delta)}{d|\Delta|} = \frac{d}{d|\Delta|}\ln[g(\Delta)]$.

The first term ($\ln q$) reflects the contribution of $U_0(n)$. Since it is a constant, decreasing impatience requires the second term to increase with $n$.

To clarify the role $\Delta$ plays here, suppose for a moment that $g\big(\Delta(n)\big) = e^{-k_2|\Delta(n)|}$ for negative $\Delta$. In this case, $h(\Delta)$ becomes a negative constant. Nevertheless, because $\frac{d|\Delta|}{dn}$ decreases with $n$ (See Figure 3 (left)), $\frac{U'(n)}{U(n)}$ still increases with $n$. Thus, $\Delta$ relaxes the conditions required for decreasing impatience: a slower-than-exponential decay of $g$ becomes sufficient but not necessary. This is illustrated by Figure A1 in Appendix A1, which shows that $U(n) = e^{-k_2|\Delta(n)|}$ still produces decreasing impatience.

If we instead use the hyperbolic form $g\big(\Delta(n)\big) = \frac{1}{1+k_2|\Delta(\mathrm{n})|}$ for negative $\Delta$ as proposed in eq. (7), eq. (11) becomes

$$\frac{U'(n)}{U(n)} = \ln q - \frac{k_2}{1+k_2|\Delta|}\frac{d|\Delta|}{dn}. \tag{12}$$

Figure 3 (left) shows that under this form of $g$, $|\Delta|$ increases approximately linearly with $n$ such that $|\Delta(n)| \approx Cn$ for small $n$, where $C > 0$. Substituting this into $U(n)$ yields near hyperbolic discounting. For large $n$, $|\Delta|$ saturates ($\frac{d|\Delta|}{dn} \approx 0$) such that $\frac{U'(n)}{U(n)}$ approaches a constant, resulting in near exponential discounting. This is illustrated in Figure 3 (right), showing that $U$ is approximately hyperbolic for short delay $n$ but deviates as $n$ grows. A detailed mathematical analysis is provided in Appendix A1. Furthermore, the delay scale ($n_{\text{transition}}$) at which the transition from hyperbolic to exponential discounting occurs is governed by the hazard probability $p$ (where $n_{\text{transition}} \sim 1/p$; see Appendix A1). If the DM assumes a small $p$, $|\Delta(n)|$ remains linear over a broad range of $n$, causing the DM's behavior to resemble pure hyperbolic discounting, as empirically observed in Myerson & Green (1995). Conversely, if the DM assumes a large $p$, $|\Delta(n)|$ flattens quickly, and the DM exhibits strong present bias for immediate rewards but near time consistency thereafter. This closely aligns with the quasi-hyperbolic ($\beta$–$\delta$) discounting models (Laibson, 1997) and related empirical observations (Cheung et al., 2026). The interplay between the functional form of $g$ and the negative surprise $\Delta$ induced during outcome resolution thus provides a unified account for the disparate patterns of impatience observed in behavioral studies.

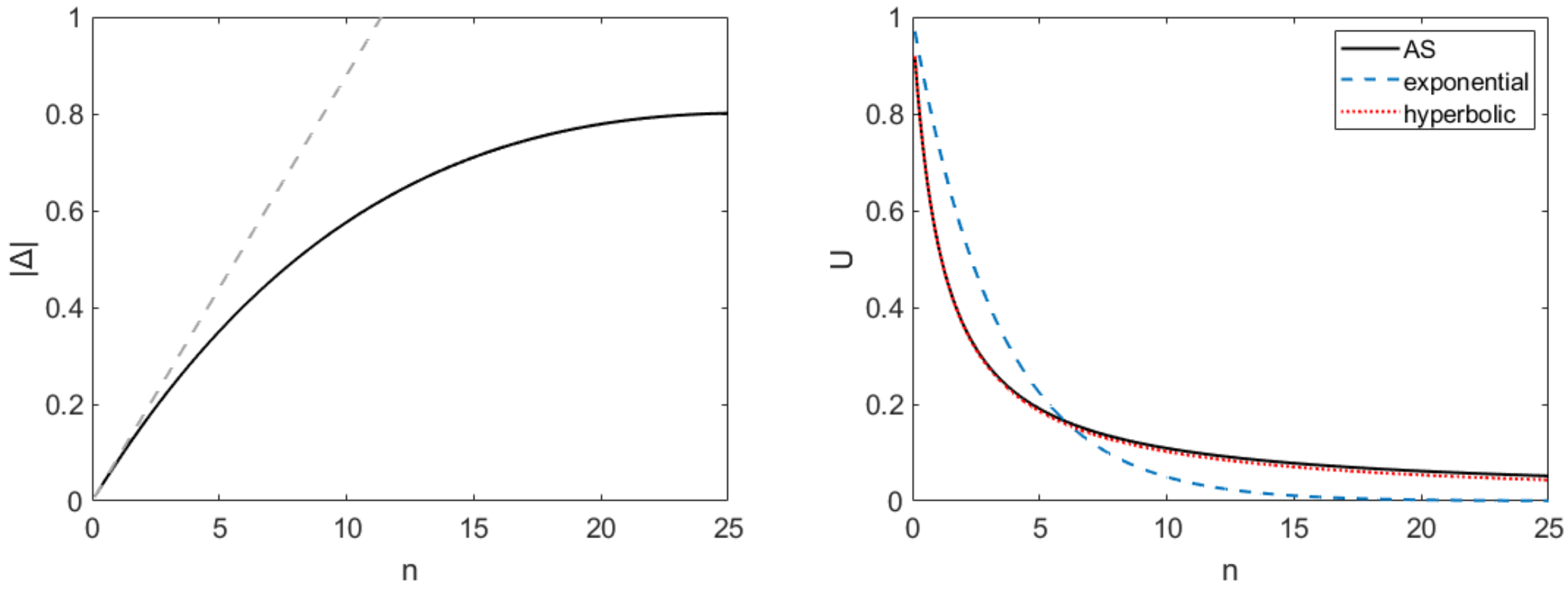

Figure 3: Left: The change in the magnitude of the surprise $|\Delta|$ when delay $n$ varies (Solid black line). The dashed gray line is a linear plot with a slope of 0.088, included to aid comparison. Right: The discount factor $U$ for the ASH model (where $g(\Delta) = \frac{1}{1+k_2|\Delta|}$), exponential discounting ($U = e^{-k_{\exp} n}$) and hyperbolic discounting ($U = \frac{1}{1+k_{\text{hypo}} n}$). $k_{\exp} = 0.3$, $k_{\text{hypo}} = 0.88$, $k_2 = 10$, $k = 3$, $\alpha = 1.6$, $p = 0.03$ for both plots. $U$ closely aligns with the hyperbolic discount curves for small $n$ before deviating as $n$ increases.

## 4.2 Aversion to timing risk

So far, we have been focusing on fixed reward delay. In many decision-making contexts, however, the timing of a delayed reward is uncertain—a condition referred to as timing risk (also called delay risk in Ebert (2021) or time lottery in Dejarnette et al. (2020)). For instance, the reward may be delivered at $t = n - 1$ with probability $p_{\text{tr}}$ or at $t = n + 1$ with probability $1 - p_{\text{tr}}$.

Expected utility theories predict that an option with timing risk is preferred to one with a deterministic delay in reward delivery equal to the corresponding expected time. Since any decreasing and bounded discount factor is necessarily convex, Jensen's inequality implies that the probability-weighted utility across uncertain delivery times is larger than the utility at the expected time. Although alternative averaging schemes have been proposed (Berman & Kirstein, 2021), they predict timing-risk neutrality at best, which is inconsistent with empirical findings showing general timing-risk aversion in DMs (Dejarnette et al., 2020; Onay & Öncüler, 2007). Moreover, this discrepancy cannot be resolved by expected utility-based hazard models (Myerson & Green, 1995; Sozou, 1998), as potential early delivery reduces total exposure to hazard, which ceases upon reward delivery, thereby increasing the overall expected value (see Appendix A2 for the mathematical details).

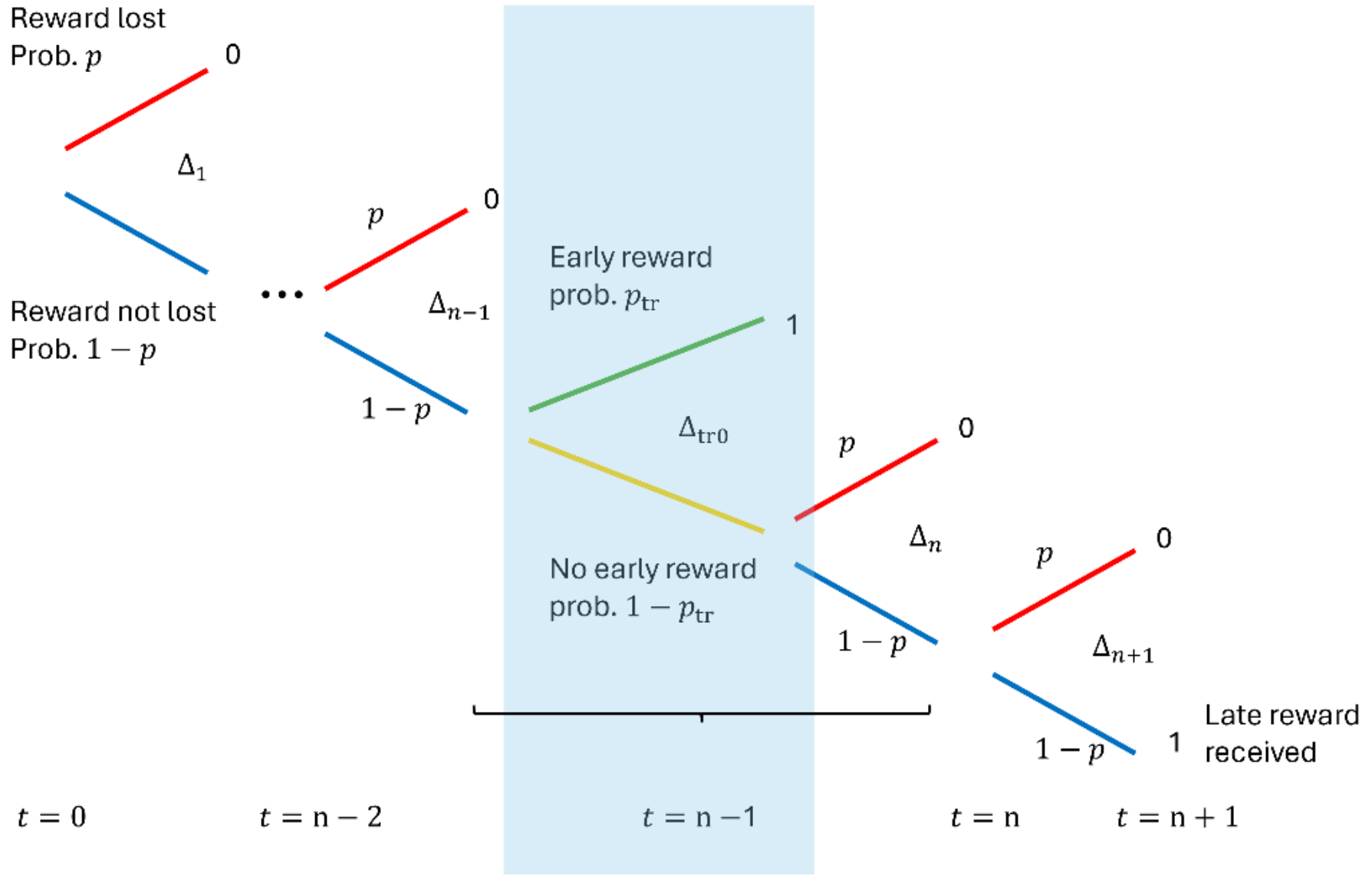

Figure 4: The branching scheme for an option with timing risk. The highlighted part illustrates the resolution of reward timing, which gives rise to timing-risk aversion in the ASH model.

In the ASH model, resolving the uncertainty about reward timing introduces an additional source of surprise. Figure 4 illustrates this using the branching scheme for an option with timing risk. At $t = n - 1$, aside from the standard branches resolving hazard occurrence (the blue-red branches), an extra set of branches resolves reward timing (the yellow-green branches). These 'reward-timing branches' mirror the structure of a probabilistic binary gamble: early (late) reward delivery represents the favorable (unfavorable) outcome owing to differing hazard exposure durations. We can therefore infer from the standard risk preferences in binary gambles that this timing-risk-induced surprise ($\Delta_{tr0}$) is negative except when $p_{tr}$ is very small. The total surprise ($\Delta_{tr}$) can be expressed as the sum of timing-risk component $\Delta_{tr0}$ and a hazard-risk component ($\Delta_{haz,tr} = \sum_{t=1}^{n+1} \Delta_t$). Since these two types of risk involved are conceptually different, we assume that they may be weighed differently (see Section 5.3 for more discussion), as shown in eq. (13).

$$\Delta_{tr} = \Delta_{haz,tr} + K_{tr}\Delta_{tr0}, \tag{13}$$

where $K_{tr}$ is the weighting factor. Please refer to Appendix A2 for the full mathematical derivation of $\Delta_{haz,tr}$ and $\Delta_{tr0}$. If $K_{tr}$ is sufficiently large, the negative $\Delta_{tr0}$ would provide enough contribution to drive timing-risk aversion.

To illustrate the resulting behavioral patterns under timing risk, Figure 5 compares the discount factor for the option with timing risk as depicted in Figure 4 ($U_{tr}$) with an option that has fixed reward delay at $t = t_{fix}$ ($U_{fix}$), where $t_{fix} = p_{tr}(n-1) + (1-p_{tr})(n+1)$, using the $g$ in eq. (7). Figure 5 (left) shows that when $\Delta_{tr0}$ is omitted, leaving only hazard-induced surprise (i.e. $K_{tr} = 0$, red line), the model predicts less discounting for the timing-risk option, in line with expected utility-based hazard models. However, when $\Delta_{tr0}$ is included ($K_{tr} = 1$, dashed black line), the bias for timing-risk vanishes. If the DM weighs the timing risk more heavily (e.g. because the resolution of reward timing occurs over a much shorter timeframe compared to the ongoing hazard exposure, drawing more attention), prominent timing-risk aversion can be reproduced ($K_{\mathrm{tr}} = 10$, solid black line). Timing-risk aversion is stronger when the average delay is short (small $n$), consistent with empirical observations (Ikink et al., 2024). This can be explained intuitively by the fact that $\Delta_{\mathrm{tr0}}$ does not scale with $n$ (Appendix A2), such that its relative contribution to the utility is diluted by hazard-driven surprise when $n$ is large for a fixed $K_{\mathrm{tr}}$.

Figure 5 (right) shows that the relationship between timing-risk aversion and $p_{\mathrm{tr}}$ resembles the risk preference patterns observed in probabilistic binary gambles when $\Delta_{\mathrm{tr0}}$ is heavily weighed (solid black line). This is unsurprising, since, as discussed above, resolving reward timing is conceptually analogous to evaluating binary gambles in the ASH model. This dependence is also consistent with experimental findings showing that timing-risk aversion is reduced or even reversed for very small $p_{\mathrm{tr}}$ (Onay & Öncüler, 2007), or more generally, for negatively skewed distributions of reward timing (Ebert, 2021).

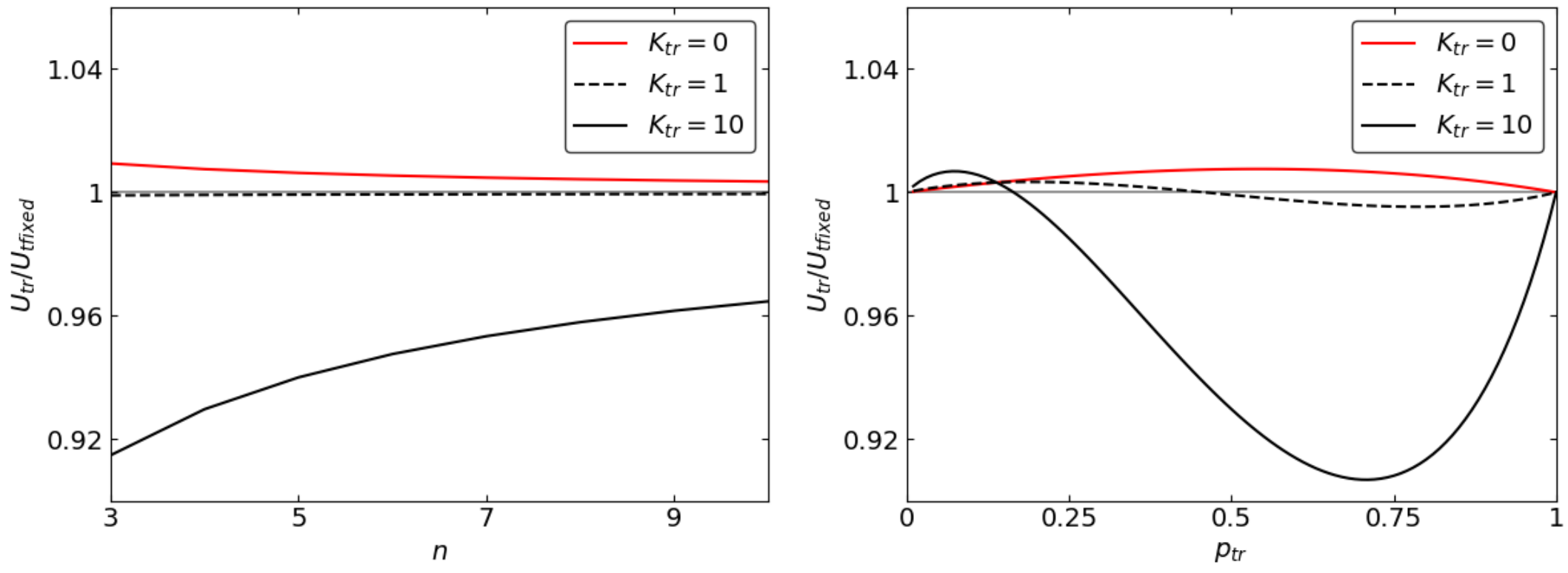

Figure 5: The ratio of discount factors between an option with timing risk as depicted in Figure 4 ($U_{\mathrm{tr}}$) and an option with fixed reward delay at $t = t_{\mathrm{fix}}$ ($U_{\mathrm{fix}}$), where $t_{\mathrm{fix}} = p_{\mathrm{tr}}(n-1) + (1 - p_{\mathrm{tr}})(n+1)$. Lower ratios indicate stronger timing-risk aversion. The red, dashed black, and solid black lines compute $U_{\mathrm{tr}}$ when timing-risk-induced surprise $\Delta_{\mathrm{tr0}}$ is omitted ($K_{\mathrm{tr}} = 0$), included ($K_{\mathrm{tr}} = 1$) and emphasized ($K_{\mathrm{tr}} = 10$), respectively. The grey line plots $\frac{U_{\mathrm{tr}}}{U_{\mathrm{fix}}} = 1$ for reference. Left: The discount factor ratio versus average delay time $n$, with $p_{\mathrm{tr}} = 0.5$. Right: The discount factor ratio versus $p_{\mathrm{tr}}$, with $n = 4$. Other parameters are the same as those used in Figure 3.

In the analysis above, the timings of the early and late rewards are held constant, such that the interval between them is fixed. A natural question is how varying this time interval alters the model predictions. Expanding this interval exerts two distinct effects on $\Delta_{\mathrm{tr0}}$: first, reward timing is resolved earlier, which increases the probability for surviving hazard to reach this resolution step ($p_{\mathrm{survival}}$); second, the drop in expected value upon confirming that the reward is late ($E_{\mathrm{drop}}$) increases, as the probability of encountering hazard within the interval thereafter increases. Since $\Delta_{\mathrm{tr0}} \sim p_{\mathrm{survival}} {E_{\mathrm{drop}}}^{\alpha}$ (Appendix 2), both effects yield a more negative $\Delta_{\mathrm{tr0}}$ and reduced utility, implying that the DM becomes more timing-risk averse. In other words, increasing the variance of the delay magnifies timing-risk discounting in the ASH model. Conversely, as the delay variance approaches 0 (i.e. fixed reward timing), $E_{\mathrm{drop}}$ and therefore $\Delta_{\mathrm{tr0}}$ also approach 0, ensuring that timing-risk discounting vanishes in the absence of timing risk.

To summarize, the ASH model interprets uncertainty in reward timing as a probabilistic gamble, in which late reward delivery represents the unfavorable outcome because it prolongs exposure to hazard and thereby reduces the expected outcome. This interpretation explains both general timing-risk aversion and the reversal to timing-risk seeking under negatively skewed timing distributions by drawing parallels to risk preferences in probabilistic gambles.

## 4.3 Opposing effects of probabilistic risk on temporal discounting

We next examine scenarios in which a delayed reward is promised to be delivered with an explicitly stated probability $p_{\mathrm{pr}}$. We refer to this combination as dual risk, as it involves both an (explicit) probabilistic risk and hazard risk. Examples of choice problems involving pure risk and dual risk are given in Table 1. Existing experimental studies do not seem to have a consensus regarding how probabilistic risk alters temporal discounting: Some reported increased temporal discounting (Blackburn & El-Deredy, 2013; Öncüler, 2000; Y. Sun & Li, 2010) while others reported a reduction (Keren & Roelofsma, 1995; Stevenson, 1992; Ventre et al., 2025). We will show how the ASH model reconciles these conflicting findings through distinct mental representations of the outcome resolution process.

| | |
|---|---|
| **Problem 1 (involving only hazard risk)** | |
| Option 1A: gain reward immediately (riskless) | Option 1B: gain reward after delay $t$ (hazard risk only) |
| **Problem 2 (involving dual risk in correspondence to problem 1)** | |
| Option 2A: gain reward $x$ with probability $p_{\mathrm{pr}}$ (probabilistic risk only) | Option 2B: gain reward after delay $t$ with probability $p_{\mathrm{pr}}$ (dual risk) |

Table 1: Example of a problem with an option involving only hazard risk (problem 1) and its corresponding problem with an option involving dual risk (problem 2).

In the ASH model, option evaluation requires creating mental pathways for the process of outcome resolution. While this may be straightforward for options with simple structures, those with complicated structures usually offer some freedom in how this could be performed. For options with dual risk, one such freedom involves how the (explicit) probabilistic risk interacts with the (implicitly assumed) hazard risk. More specifically, since hazard risk is ‘probabilistic’ in nature in the model, it is unclear whether the (explicit) probabilistic risk should be separately resolved, as exemplified by the branching scheme in Figure 6a (hereafter Scheme A, with its variants also analyzed later), or be incorporated into the hazard by increasing the hazard rate to reflect the expected reward reduction (associated with the explicit risk), as depicted by Figure 6b (Scheme B).

In real life, there are often contexts that bias which interpretation is adopted by DMs. For instance, it makes sense to incorporate the risk of heart attack into the hazard while keeping the risk of a failed surgery separated. However, such contexts are lacking in most laboratory studies, and how risks are resolved was unknown to the subjects. Which interpretation they take may then come down to other factors, such as procedural bias, framing and cognitive limitations of DMs. Here, we hypothesize that DMs simplify the mental simulation of outcome resolution in order to reduce cognitive load when facing mentally demanding tasks by incorporating the explicit risk into hazard. The experiments in Blackburn & El-Deredy (2013), Öncüler (2000) and Y. Sun & Li (2010) involve pricing an option, which is more mentally demanding than simpler preference elicitation methods, such as giving valence ratings (Stevenson, 1992) or choosing between two options (Keren & Roelofsma, 1995; Ventre et al., 2025). Under the above-mentioned assumption, DMs in experiments that use pricing to elicit preference are more likely to use Scheme B for option evaluation, while those in the other experiments are more likely to use Scheme A.

(a)

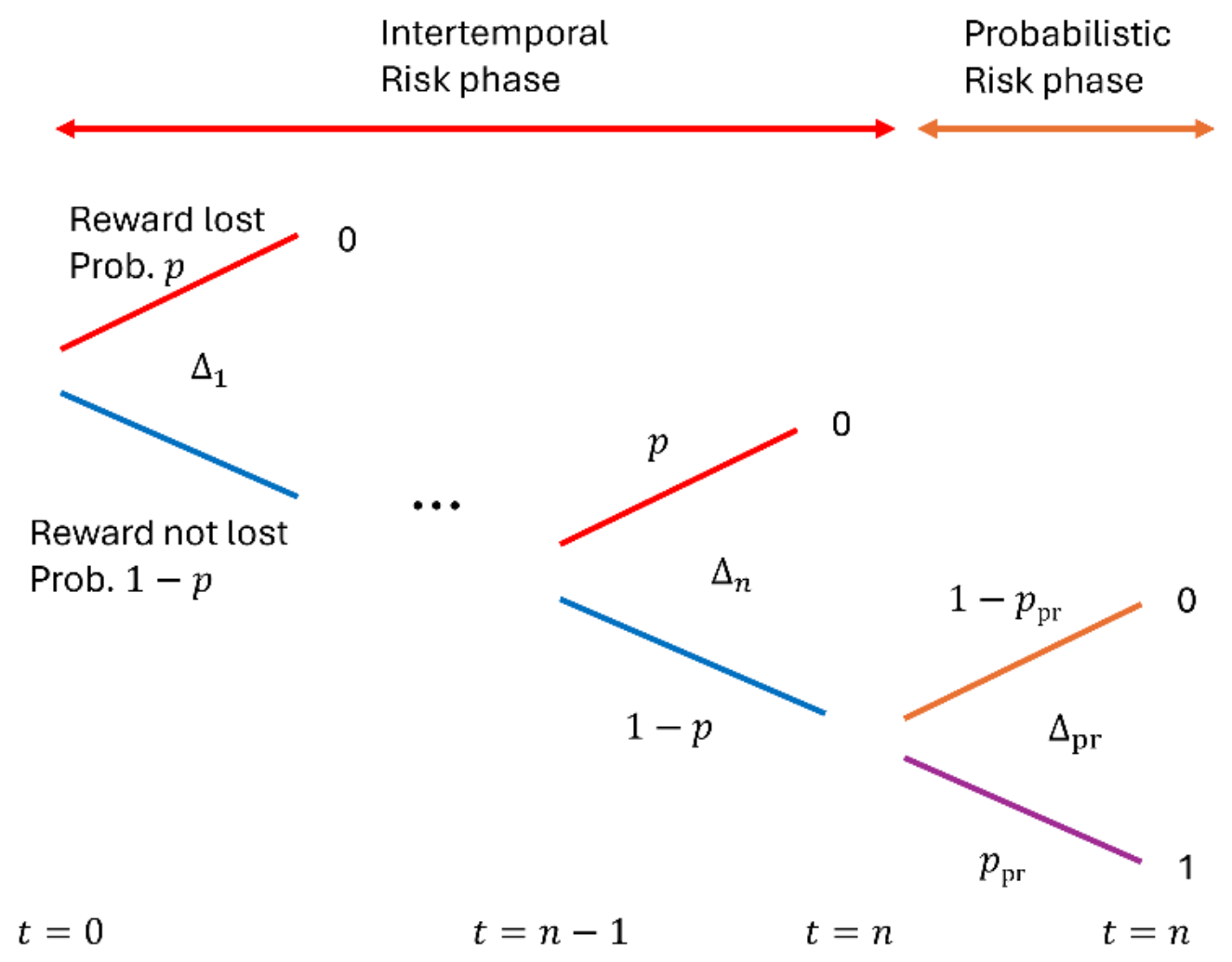

(b)

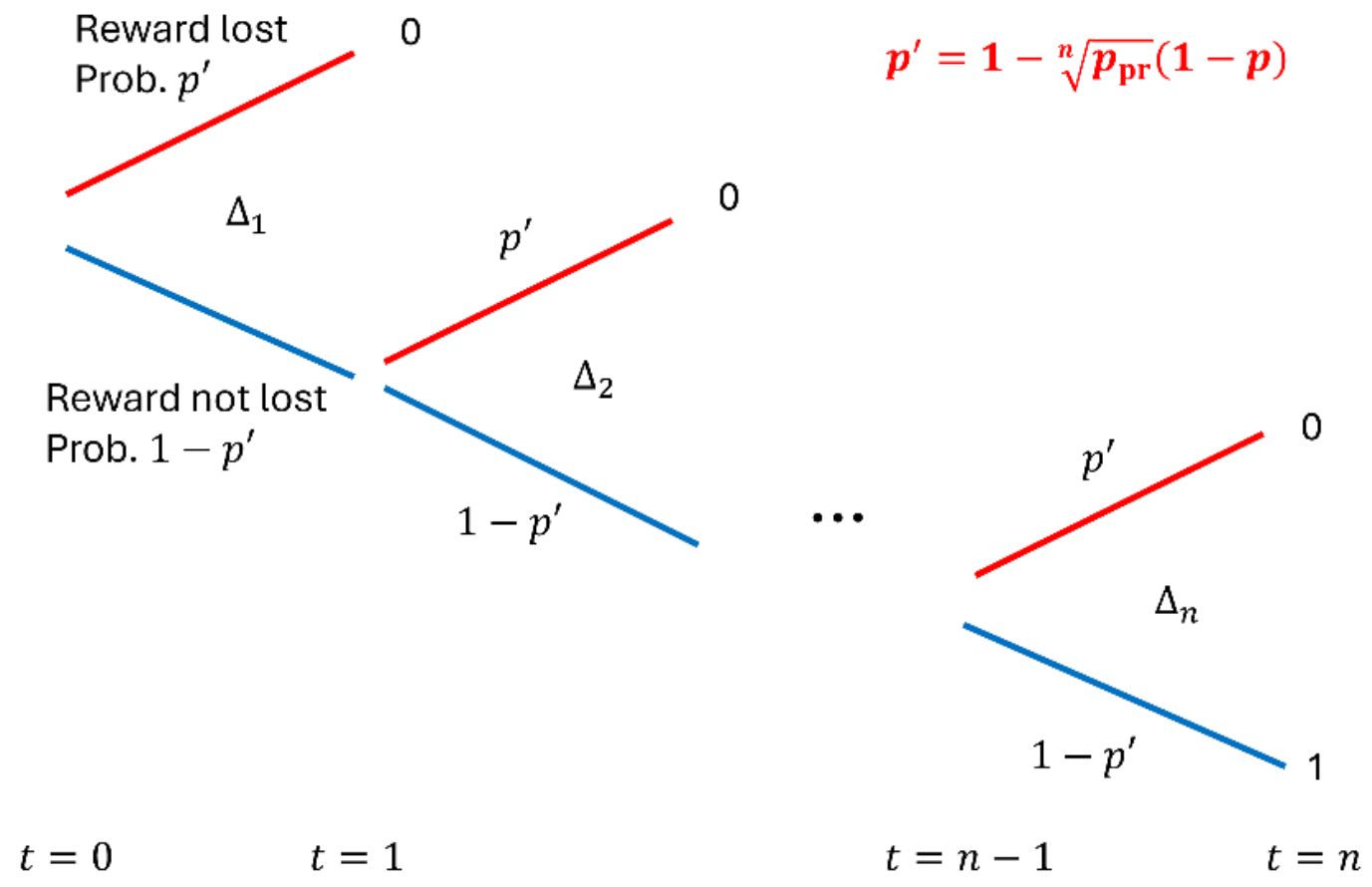

Figure 6: Possible branching schemes for a dual-risk option. (a) Scheme A: The probabilistic risk and hazard risk are resolved separately. Here, the probabilistic risk (orange-purple lines) is resolved after the hazard risk (red-blue lines). Note that it is also reasonable to resolve the hazard risk first instead (not shown). (b) Scheme B: The probabilistic risk is incorporated into the hazard by inflating the hazard probability in each time step to $1 - \sqrt[n]{p_{\mathrm{pr}}}(1-p)$.

To understand whether and how these schemes affect the influence of probabilistic risk on temporal discounting in the ASH model, we consider the problems in Table 1 and compare the discounting of the dual-risk option ($U_{\mathrm{pt}}$) (option 2B) relative to its probabilistic risk-only counterpart ($U_{\mathrm{p}}$) (option 2A) against the discounting of the hazard risk-only option ($U_{\mathrm{t}}$) (option 1B) relative to the certain reward ($U_{\mathrm{certain}}$) (option 1A) for each scheme. This can be expressed mathematically by the 'discount ratio' $D$:

$$D = \frac{\frac{U_{\mathrm{pt}}}{U_{\mathrm{p}}}}{\frac{U_{\mathrm{t}}}{U_{\mathrm{certain}}}} = \frac{U_{\mathrm{pt}}}{U_{\mathrm{p}} U_{\mathrm{t}}}, \tag{14}$$

where $D > 1$ ($D < 1$) means that probabilistic risk reduces (increases) the discounting associated with hazard risk. Please note that $U_{\mathrm{pt}}$ differs between Scheme A and Scheme B due to their distinct branching structure.

As in our analysis for timing risk, we assume the surprise associated with hazard risk and probabilistic risk is weighted differently, with weighting factors $K_{\mathrm{haz}}$ and $K_{\mathrm{pr}}$ respectively (see eq. (5) and Section 5.3 for discussion). We also exclude the regime of very small $p_{\mathrm{pr}}$ or large $np$, since the value of $D$ is largely behaviorally irrelevant there: the large hazard-induced reduction in the expected reward makes the utility of all options negligibly small, so that they are evaluated as essentially worthless and comparison between them becomes practically meaningless.

One key difference between Scheme A and Scheme B lies in the relative contribution of their constituent risk components to surprise. In Scheme A, the temporal (hazard) and probabilistic risk components in the dual risk option remain separate, mirroring standalone hazard and probabilistic risk in the pure-risk options and their respective weighting towards surprise. In Scheme B, however, the probabilistic risk in the dual-risk option is absorbed into the hazard, eliminating the probabilistic component while inflating the hazard component of surprise, thus creating an asymmetry in scaling with the probabilistic risk-only option. This structural difference is reflected in the mathematical expression of the discount factors for Scheme A ($D_A$) and Scheme B ($D_B$), derived using eq. (1)-(6):

$$D_{\mathrm{A}} = \frac{g\left(K_{\mathrm{pr}} q^n \Delta_{\mathrm{pr}} + K_{\mathrm{haz}} {p_{\mathrm{pr}}}^{\alpha} \Delta_{\mathrm{haz}}(p)\right)}{g\left(K_{\mathrm{pr}} \Delta_{\mathrm{pr}}\right) g\left(K_{\mathrm{haz}} \Delta_{\mathrm{haz}}(p)\right)} \tag{15}$$

$$D_{\mathrm{B}} = \frac{g\left(K_{\mathrm{haz}} \Delta_{\mathrm{haz}}(p')\right)}{g\left(K_{\mathrm{pr}} \Delta_{\mathrm{pr}}\right) g\left(K_{\mathrm{haz}} \Delta_{\mathrm{haz}}(p)\right)}, \tag{16}$$

where $\Delta_{\text{pr}} = p_{\text{pr}}(1 - p_{\text{pr}})^{\alpha} - k(1 - p_{\text{pr}}){p_{\text{pr}}}^{\alpha}$, $\Delta_{\text{haz}}(x) = -C(x)(1 - x)^{n+\alpha'} \frac{1-(1-x)^{n\alpha'}}{1-(1-x)^{\alpha'}}$ with $C(x) = kx - x^{\alpha}(1 - x)^{1-\alpha}$, $p' = 1 - \sqrt[n]{p_{\text{pr}}}(1 - p)$. Please refer to Appendix A3 for the complete mathematical derivation.

Since negative surprise reduces utility, $g(\Delta)$ is decreasing with $|\Delta|$ (See eq. (6)). Hence,

$$D_{\text{A}} > \frac{g(K_{\text{pr}}\Delta_{\text{pr}} + K_{\text{haz}}\Delta_{\text{haz}}(p))}{g(K_{\text{pr}}\Delta_{\text{pr}})g(K_{\text{haz}}\Delta_{\text{haz}}(p))} . \tag{17}$$

By taking the logarithm of $D_{\text{A}}$ and applying Jensen's inequality, it follows from eq. (17) that a sufficient condition for $D_{\text{A}} > 1$ is that $-\ln(g(\Delta))$ is concave, or equivalently, that $h(\Delta)$ is non-decreasing with $|\Delta|$ (as a reminder, $h(\Delta) = \frac{1}{g(\Delta)} \frac{dg(\Delta)}{d|\Delta|}$), regardless of the value of $K_{\text{haz}}$ and $K_{\text{pr}}$. The functional form of $g$ in eq. (7) satisfies this condition (See eq. (12) and Appendix A3).

$D_{\text{B}}$ behaves differently. For the dual-risk option in Scheme B, because the probabilistic risk is embedded within the inflated hazard rate $p'$, its contribution to surprise is scaled by $K_{\text{haz}}$ along with the 'baseline' hazard risk. To study the effect of this scaling, we take the log-derivative of eq. (16) with respect to $K_{\text{haz}}$:

$$\frac{d}{dK_{\text{haz}}} \ln(D_{\text{B}}) = |\Delta_{\text{haz}}(p')| h(K_{\text{haz}}\Delta_{\text{haz}}(p')) - |\Delta_{\text{haz}}(p)| h(K_{\text{haz}}\Delta_{\text{haz}}(p)). \tag{18}$$

If $\frac{d}{dK_{\text{haz}}} \ln(D_{\text{B}}) < 0$, then $D_{\text{B}}$ decreases with $K_{\text{haz}}$, such that $D_{\text{B}} < 1$ might occur when $K_{\text{haz}}$ is sufficiently large. Since $p' > p$, eq. (18) implies that $\frac{d}{dK_{\text{haz}}} \ln(D_{\text{B}}) < 0$ if $|\Delta_{\text{haz}}(\text{x})| h(\Delta_{\text{haz}}(x))$ decreases with $x$.

In the regime of interest (non-small $p_{\text{pr}}$ and small $np$), $|\Delta_{\text{haz}}(x)|$ is increasing (see Figure 3 and Appendix A1). Moreover, as previously analyzed, an increasing $h(x)$ (i.e. becoming less negative) contributes to decreasing impatience (see eq. (11)). Thus, the condition that $|\Delta_{\text{haz}}(\text{x})| h(\Delta_{\text{haz}}(x))$ decreases with $x$ (i.e. becoming more negative) can be understood intuitively: decreasing impatience through $h$ must not be so strong that it dominates the growth in $|\Delta_{\text{haz}}(\text{x})|$. The functional form of $g$ in eq. (7) satisfies this condition. In particular, for this $g$, $D_{\text{B}} < 1$ if $|\Delta_{\text{haz}}(p')| > |\Delta_{\text{haz}}(p)|(1 + k_{\text{eff}}|\Delta_{\text{pr}}|) + \frac{|\Delta_{\text{pr}}|}{R}$, where $R = \frac{K_{\text{haz}}}{K_{\text{pr}}}$ (see Appendix A3).

Figure 7 plots $D_{\text{A}}$ and $D_{\text{B}}$ in eq. (15) and (16) using the $g$ in eq. (7) and $R = 3$. As in our analysis, $D > 1$ (solid black line), when the probabilistic risk is separate (Scheme A), indicating a reduced discount; while incorporating the probabilistic risk into the hazard (Scheme B) results in $D < 1$ (dashed line), indicating an increased discount. This is consistent with empirical behaviors: subjects in pricing experiments (Blackburn & El-Deredy, 2013; Öncüler, 2000; Y. Sun & Li, 2010), in which we hypothesize that Scheme B would be used, exhibit higher temporal discounting when probabilistic risk is added, while subjects in other experiments exhibit reduced discounting (Keren & Roelofsma, 1995; Stevenson, 1992; Ventre et al., 2025).

Experimental studies show that the timing of probabilistic-risk resolution affects the evaluation of dual-risk (Kemel & Paraschiv, 2023). One may ask whether $D_\mathrm{A}$ remains larger than 1 if the probabilistic risk is resolved before the hazard risk instead. Since no matter when the probabilistic risk is resolved, the surprise induced by both types of risk is attenuated either by the reduced expected value or the reduced survival probability (Appendix A3), the convexity of discount curve still ensures that $D_\mathrm{A} > 1$. This is confirmed in Figure 7, which shows that despite quantitative changes, $D_\mathrm{A} > 1$ still holds when the probabilistic risk is resolved early (blue line).

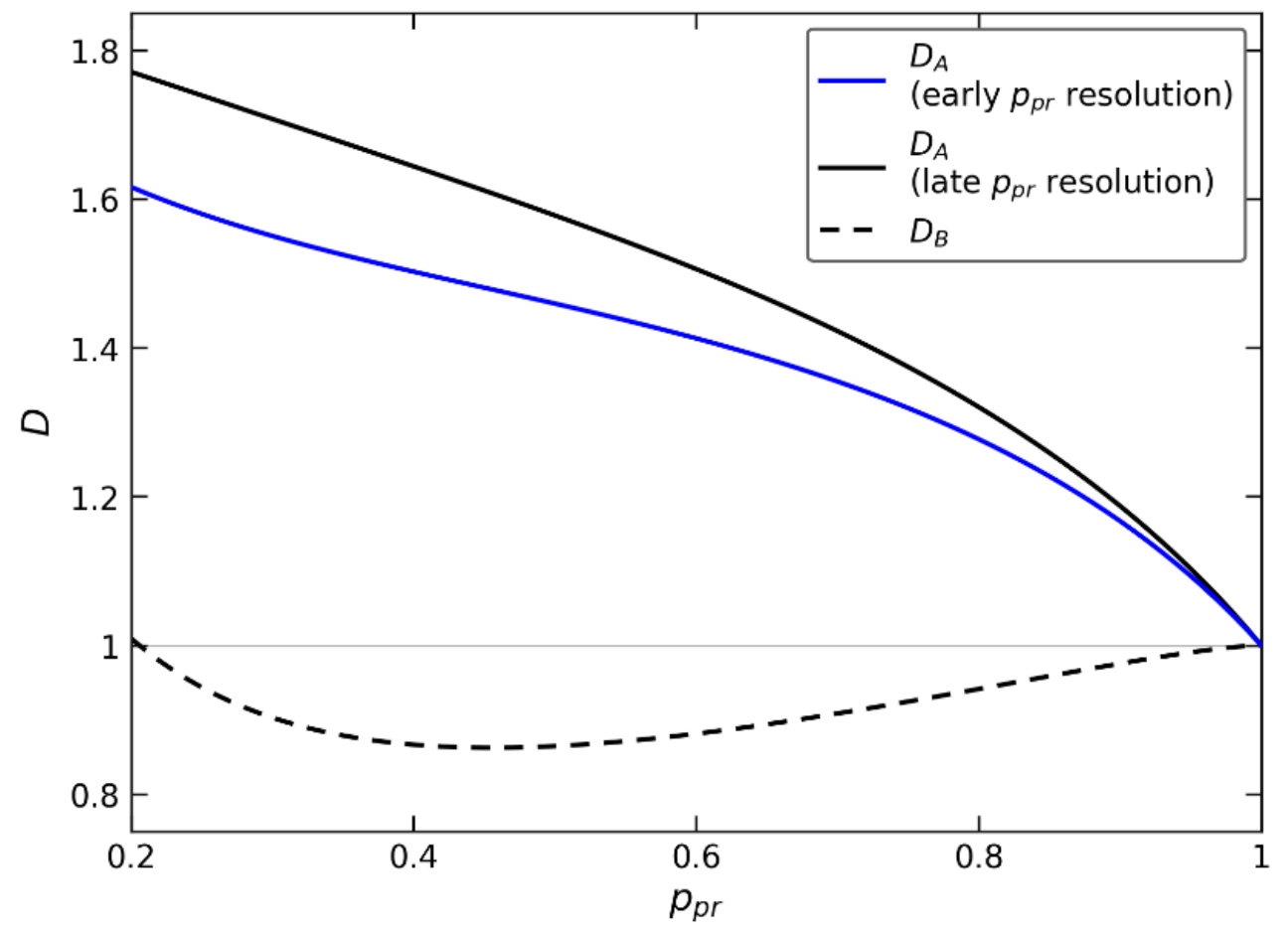

Figure 7: The discount factors for Scheme A ($D_\mathrm{A}$) and Scheme B ($D_\mathrm{B}$) in the ASH model. The solid black (blue) line plots $D_\mathrm{A}$ with the probabilistic risk resolved after (before) the hazard risk, and the dashed line plots $D_\mathrm{B}$. A larger (smaller) than unity $D$ suggests reduced (increased) temporal discount when probabilistic risk is present. The gray line plots $D = 1$ for reference. $k_\mathrm{eff} = 1$, $R = 3$ and $n = 4$. Other parameters of the model are the same as Figure 3 and 5.

One may also question the hypothesis of attributing differences in option evaluation to experimental methodologies. It has been shown that people evaluate the same option differently in pricing and in choice experiments when making intertemporal decisions (Kvam et al., 2023). While it does not prove that DMs use the exact representation proposed here, it supports the notion that evaluation processes may depend on different elicitation methods.

To summarize, we hypothesize that even for the same choice problem, DMs may adopt different pathways to mentally resolve an option's outcome, each producing a distinct evaluation. This allows the ASH model to reconcile the opposing empirical findings in dual-risk discounting: the observation that probabilistic risk can both attenuate and enhance temporal discounting is not contradictory, but rather reflects different mental representations of complex options in DMs, driven by factors such as preference elicitation methods.

# 5. Discussion

In this work, we extend the AS framework developed in Chan & Toyoizumi (2024) to jointly accommodate probabilistic risk and delay by treating delay as 'hazard risk' arisen from continuous exposure to events that cause reward loss. This interpretation allows intertemporal choices to be represented as compound lotteries of probabilistic gambles, with utility computed by aggregating the anticipated surprise generated along each gamble. The resulting ASH model not only reproduces key empirical phenomena in intertemporal decision-making, including time inconsistency, timing-risk aversion and dual-risk interaction, but also reconciles the disparate or seemingly contradictory empirical findings and theories in these behaviors under a unified framework.

Our work highlights that changes in DMs' mental representation of the outcome resolution process fundamentally alter how uncertainty is construed, leading to distinct choice behaviors. By elucidating how such mental representations are shaped by contextual cues, attentional salience and cognitive constraints, our framework could provide a foundation for modeling broader behavioral biases, such as framing effects and other context-dependent anomalies.

## 5.1 Relationship between the ASH model and existing theoretical frameworks and models

### *5.1.1 Theoretical framework modeling delay as hazard*

Several previous works assume potential reward loss due to hazard in intertemporal choice, as in this work. Sozou (1998) and Myerson & Green (1995) explained decreasing impatience by using a stochastic hazard rate or one that increases with time. By contrast, the ASH model reproduces decreasing impatience even with a constant hazard rate. This difference arises because those models are under an expected utility framework, whereas both the anticipated-surprise function (eq. (2)) and the utility function (eq. (7)) in the AS framework are convex. Halevy (2008) showed that a sufficient condition for decreasing impatience for hazard-based models is a convex transformation of total hazard, and the utility for an option with simple hazard risk in the ASH model (eq. (9)) satisfies this condition.

While both Halevy's framework and the ASH model account for decreasing impatience through similar mechanisms, the ASH model additionally captures the interactions among multiple types of risk. This allows it to address a broader range of choice problems, whereas Halevy's descriptive theoretical framework has a narrower scope, focusing primarily on simple intertemporal choices with fixed delays.

### *5.1.2 Theoretical framework modeling probabilistic risk as delay*

While hazard-based approaches and the ASH model map temporal delay onto probabilistic risk, an alternative line of research—spearheaded by Rachlin et al. (1991)—links the two in the opposite direction by transforming probabilistic risk into delay. As described in the introduction, this approach assumes that DMs repeatedly sample probabilistic outcomes separated by inter-trial delays. The AS framework similarly assumes mental simulation across distinct outcome resolution pathways, but applies non-linearity to outcome value via a convex surprise function; whereas Rachlin's model applies non-linearity to outcome probability via a hyperbolic transformation of the inter-trial delay. Consequently, unlike the AS framework, Rachlin's framework cannot predict risk seeking behavior for small-probability gains.

Furthermore, it is unclear how options containing more than two possible outcomes are evaluated under Rachlin's framework. More specifically, substantial freedom remains regarding whether delay and outcome value are sampled separately for each outcome path or interactively, and in case of separate sampling, how the resulting effective delays across paths are combined for option evaluation. By contrast, the AS framework assumes that surprise accumulates additively across pathways, and this minimal assumption alone is sufficient to account for a wide range of empirical phenomena.

#### *5.1.3 Descriptive mathematical model of dual risk*

Descriptive mathematical models (Baucells & Heukamp, 2010; Rambaud & Pérez, 2020; Vanderveldt et al., 2015; Yi et al., 2006) bypass the cognitive mechanisms underlying the evaluation of risk and time. Instead, they proposed mathematical expressions as functions of outcome probability and delay, with the primary aim of fitting experimental data. While such models can sometimes offer valuable insights into the potential relationship between the two types of risk and the behavioral anomalies that arise from them (Luckman et al., 2018), their scope is restricted to options with simple resolution structures. Applying these models to options with higher-order uncertainties (e.g. compound lotteries or timing risk) requires either averaging out the high-order uncertainties before computing the utility or averaging the utilities across the final outcomes in each resolution pathway. Both approaches are problematic: the former eliminates the higher-order effects (e.g. Ellsberg's paradox and blackjack problems; (Chan & Toyoizumi, 2024)), while the latter potentially yields incorrect predictions, such as failing to predict timing-risk aversion.

The AS framework, by comparison, explicitly models the mental pathways of outcome resolution that reflect the option structure, enabling it to reproduce complex choice behaviors. For example, the empirically observed opposing effects of probabilistic risk on temporal discounting are accounted for by distinct mental representations of explicit probabilistic risk in dual-risk options. On the other hand, descriptive models, lacking a mechanism for alternative risk representations, can only fit one directional effect or the other.

### 5.2 Potential issues in the ASH model and workarounds

#### *5.2.1 Outcome scaling and magnitude effects for probabilistic risk*

As mentioned in Section 2, eq. (6) and (7) produce problematic predictions in both the loss and mixed domains. In the loss domain, where the expected value $U_0 < 0$, the negative sign in $U_0$ causes the surprise $\Delta$ to have a reverse effect: positive surprise reduces the utility $U$, and negative surprise increases it. This violates the empirically observed reflection effects (Kahneman &

Tversky, 1979). In the mixed domain where $U_0 \approx 0$, $\Delta$ becomes ineffective at modulating $U$. In the extreme case of $E = 0$, the model yields $U = 0$ for all $\Delta$, predicting risk neutrality. This contradicts empirical findings by Tom et al. (2007) that people generally reject symmetric gambles of zero expected value, such as (-1, 0.5; 1, 0.5), which implies $U < 0$.

Another issue concerns the high sensitivity of $\Delta$ to outcome magnitude due to the convexity of the surprise function. Under large outcomes, the correction applied to $U_0$ by $\Delta$ dominates the computation of $U$ (see Appendix A4 for an illustration), producing magnitude effects far more extreme than those observed empirically in Cruz Rambaud et al. (2023). One potential workaround is to scale outcomes to the unit interval $[0,1]$ before computing utility, followed by reverse scaling of $U$ (Appendix A4). This full normalization, however, eliminates magnitude effects entirely. A partial normalization, e.g. scaling outcomes to $[0, f(x_{\min}, x_{\max})]$ where $1 < f(x_{\min}, x_{\max}) < x_{\max}$, may attenuate excessive outcome sensitivity while preserving realistic magnitude effects (Appendix A4).

Broadly speaking, it is important to clarify how affective components of the utility (e.g. anticipated surprise) interact with value components (outcome size), and how this interaction differs in the gain and loss domains. This may offer new perspectives on outcome-size and domain-dependent anomalies in decision-making under uncertainty.

### *5.2.2 Magnitude effects for delayed outcomes*

While increasing reward magnitude magnifies risk aversion under probabilistic risk, it reduces discounting from delay (Cruz Rambaud et al., 2023; Meyer, 2015). These opposing magnitude effects are often interpreted as evidence that DMs conceptualize probabilistic risk and delay as distinct constructs (Andreoni & Sprenger, 2012; Green & Myerson, 2004). Because the current ASH model directly transforms delay into probabilistic risk via hazard, it cannot reproduce both magnitude effects simultaneously.

We outline two possible explanations for the temporal magnitude effects within the AS-hazard framework. First, larger rewards may lead DMs to discretize the hazard process into fewer resolution events. In the ASH model, this corresponds to reducing the number of time steps $n$ while increasing the hazard probability $p$ in each branch (Figure 2). Because each successfully evaded hazard updates the expected reward to a higher value, encountering a hazard in subsequent branches generates a more negative surprise, thereby increasing temporal discounting. As for why such hazard discretization might occur, one possibility is that reward magnitude modulates subjective time perception (Apaydın et al., 2018; Failing & Theeuwes, 2016), which may influence how DMs segment time when evaluating delays. Second, DMs may derive positive utility from waiting for a reward (Hardisty & Weber, 2020; Iigaya et al., 2020; Loewenstein, 1987). For instance, option evaluation may involve combining this anticipatory utility and the surprise-driven utility in eq. (7) additively, and the observed magnitude effects could arise if the anticipatory utility scales superlinearly with reward magnitude.

To conclude, while the ASH model in its current form does not account for opposing magnitude effects for probabilistic risk and delay, it is not inherently incompatible with them. More generally, opposing magnitude effects alone are insufficient to rule out a possible cognitive relationship between probability and time.

## 5.3 Interactions between risk types implied in the ASH model and physiological findings

In the ASH model, options involving multiple interacting risks are decomposed into a combination of probabilistic binary gambles. A naive interpretation of this formulation is that the brain treats all risk types as identical and processes them indiscriminately through a single neural system. However, this contradicts neurophysiological findings showing that risk and delay activate distinct brain regions (Luhmann et al., 2008; Weber & Huettel, 2008).

A more plausible interpretation is that distinct neural populations are recruited to encode different types of risks by representing them as sequences of risk-resolution events. The neural activity reflecting this encoding is then integrated downstream for the computation of anticipated surprise. The resulting anticipated surprise signal is subsequently transmitted further downstream to generate an overall utility value and determine choice preference. This modular processing paradigm justifies risk-specific scaling in utility computation (Sections 4.2 and 4.3), because inputs from different risk-encoding neural groups do not necessarily contribute equally to downstream activity. This asymmetric contribution may be intrinsic (e.g. reflecting evolutionary optimization) or induced and modulated by physiological states or external stimuli. Indeed, pharmacological findings confirm that dopamine-modulating medications can alter the ratio of temporal discounting to probabilistic risk discounting (Arrondo et al., 2015).

A central question in neuroeconomics concerns the neural substrates underlying risk encoding and evaluation. In the ASH model, all risks culminate in surprise, which measures the deviation of an updated expected outcome—or, more broadly, state value (e.g. regarding future consumption or financial well-being)—from prior expectations. This computational mechanism is highly reminiscent of reward prediction error (RPE) signaling mediated by dopaminergic pathways. While dopamine's role in evaluating probabilistic risk is well established (Wise, 2004), recent neurophysiological studies demonstrate that manipulating the dopaminergic state also alters temporal discounting (Arrondo et al., 2015; Foerde et al., 2016) and modulates temporal magnitude effects (Münte et al., 2018). These empirical findings suggest that dopamine serves as a common neurobiological substrate contributing to the evaluation of both temporal and probabilistic risks. Further experimental studies elucidating how RPE signals are integrated downstream and how utility is represented in the brain would help provide a more complete account of the neuropsychological mechanisms of utility formation and explain the key features of the AS framework, e.g. convexity of the surprise function.

## 5.4 Extension and application of the ASH framework

In this work, we employed the concept of ‘hazard’ to refer to unforeseen events that result in reward loss. More generally, however, hazard can be conceptualized as any stochastic transition yielding an unexpected positive or negative outcome. Under this broader interpretation, the ASH framework can address a wide range of decision-making problems involving complex risks and uncertainties.

### *5.4.1 Extension to the loss domain*

One natural extension of the framework involves intertemporal choices in the loss domain (i.e., delayed negative outcomes). In this context, hazard represents unexpected relief events that avert impending losses. Since small-probability positive outcomes generate positive anticipated surprise, a naïve application of the framework predicts that delayed negative outcomes are preferred over immediate ones. However, empirical studies on intertemporal loss remain scarce due to methodological limitations in representing true losses in laboratory settings (Kemel & Paraschiv, 2023), and among existing studies, conclusions are often mixed (Hardisty & Weber, 2020; H. Y. Sun et al., 2022). These mixed findings suggest that several competing mechanisms, e.g. anticipated dread (Loewenstein, 1987; H. Y. Sun et al., 2022), consumption smoothing (Jappelli & Pistaferri, 2010) and framing, might be involved in shaping the evaluation of future loss. Investigating how these mechanisms interact with hazard perception and influence DMs' mental representation of loss offers a promising direction for future research.

### *5.4.2 Practical applications*

The AS-hazard framework provides a useful foundation for analyzing real-world decision-making problems across behavioral economics, health psychology, and organizational management. We will discuss two potential applications below.

**Procrastination**

Here we consider a DM evaluating whether to commit to a task whose completion time is highly variable (e.g., dieting, an R&D project, or infrastructure development) and subject to a deadline—either soft or hard—after which a negative outcome occurs (e.g. adverse health effects, losing out to competitors, or loss of public support). This can be modeled within the ASH framework as depicted in Figure 8 (left). In this representation, hazard corresponds to task completion (a positive outcome), while the outcome at the deadline is negative. Soft deadlines are modelled as timing risk (as in Section 4.2) through additional branches resolving deadline timings (not shown in Figure 8).

From the expected-utility perspective, the DM faces a tradeoff: committing to the task immediately postpones other utility-generating actions (e.g., eating enjoyable food, allocating resources elsewhere), imposing temporal discounting on their utility $U_A$, whereas delaying commitment allows the DM to take those actions now at the cost of reducing the probability of timely task completion $U_0$. When the deadline is far away, the reduction in $U_0$ from delaying is small, so procrastination arises from discounting of $U_A$. As the deadline approaches, the reduction in $U_0$ becomes dominant, restoring urgency and inducing commitment.

Incorporating anticipated surprise introduces an additional factor in the tradeoff. Because surprise is negative and increasing in $n$ (except at very large $n$), early commitment generates larger negative surprise, biasing the DM toward delaying commitment for longer. Since anticipated surprise also drives probabilistic risk aversion and contributes to time inconsistency, the model predicts that individuals exhibiting stronger risk aversion may paradoxically miss deadlines more often, and is consistent with empirical evidence linking time inconsistency and procrastination (Steel, 2007).

Another source of procrastination arises from deadline uncertainty. If the time distribution of the deadline has a very long tail (e.g., for negative health effects or natural-disaster risk), the statistics of the outcome change little as time passes. In the extreme case of an exponential deadline distribution, utility becomes independent of time: if the DM initially evaluates procrastination as favorable, this evaluation persists indefinitely unless internal states change. This provides a perspective on why individuals may resist changing habits to improve health, or why governments may repeatedly defer infrastructure investments that mitigate natural disasters.

**Negotiation**

Figure 8 (right) illustrates a possible branching scheme for multi-stage negotiation under the ASH framework. In this representation, each 'time step' refers to a bargaining stage in the negotiation rather than literal time progression. Hazard corresponds to negotiation breakdown. It becomes more severe in later stages (with outcomes $0 = x_n < x_{n-1} < \cdots < x_1 < 1$; see Figure 8 (right)), reflecting the wasted resources and missed opportunities associated with late-stage breakdown. Surviving all hazard leads to the conclusion of an optimal agreement (positive outcome) at $t = n$. The alternative option is to skip the negotiation altogether or terminate it early, yielding an outcome inferior to the optimal agreement but superior to breakdown.

A key insight drawn from Section 4 is that anticipated surprise can hinder successful negotiation. One might expect DMs' inclination to continue negotiating to be high in its later stages due to increased goodwill or consideration of sunk costs. However, because expectations rise and hazard worsens in these late stages, anticipated surprise becomes very negative, resulting in substantial discounting of delayed outcomes. Moreover, in this regime (small $n$), the model predicts decreasing impatience. Together, these factors imply that even if a DM initially plans to complete the entire negotiation, the increasing fear of disappointment from a potential breakdown may induce premature termination and acceptance of a suboptimal agreement. We can also infer from timing-risk aversion (Section 4.2) that having a fixed negotiation schedule increases willingness to participate compared to having a loose or unspecified schedule.

Successful negotiations, such as signed peace deals, do not always guarantee their intended positive outcomes. This can be modeled using dual risk (Section 4.3). In this context, the branching scheme in Figure 6a likely captures DMs' mental representation, since events driving peace-negotiation breakdown do not typically affect whether a signed accord delivers its intended benefits (e.g., long-term compliance, economic development, or political stability). Because probabilistic risk reduces temporal discounting under this representation, the model counterintuitively predicts that ambitious high-variance, high-payoff peace deals are preferable to modest, more implementable ones. Combined with the earlier prediction of a tendency to break off negotiations prematurely or strike suboptimal agreements, this offers an alternative angle on why temporary truces are often formed and then collapse in quick succession (e.g. the Minsk Agreements), and why complex agreements that result in long-term institutional fragility and other problems across all signatories are nonetheless concluded (e.g. the Dayton Accords).

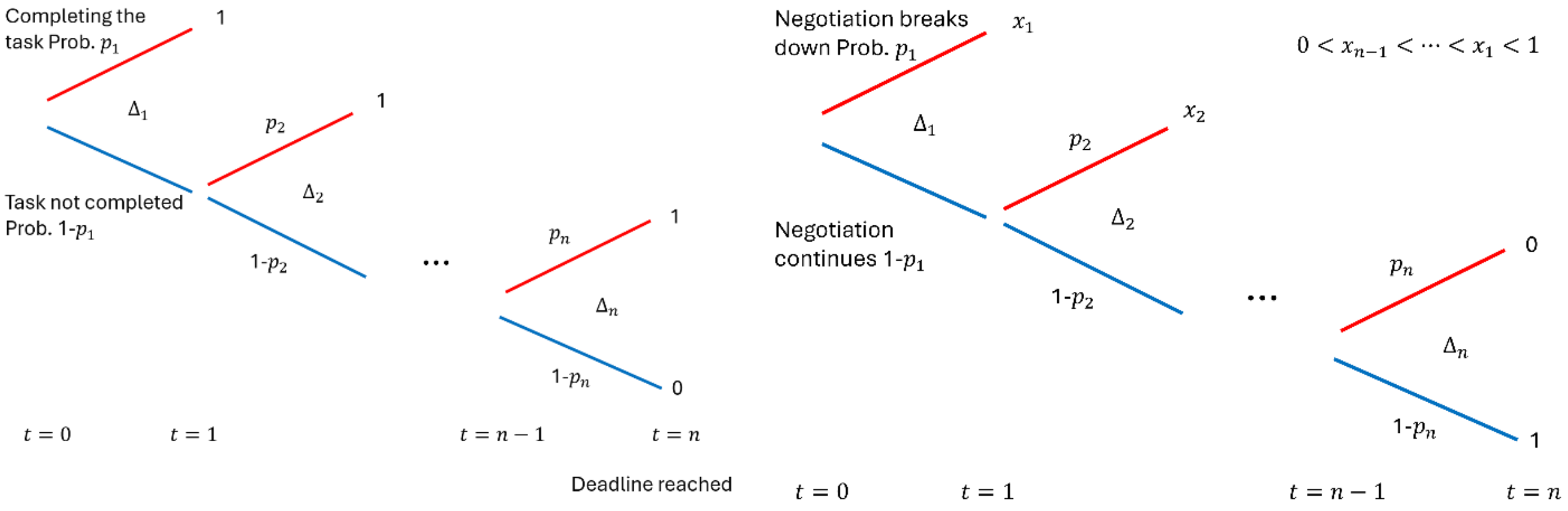

Figure 8: Branching scheme for modeling procrastination (left) and negotiation (right).

# Appendix A1: Analysis of the ASH model on time inconsistency

**The discount factor for an exponential $U(\Delta)$**

In Figure A1, we plot $U$ against $n$ assuming $U(\Delta) = e^{-k_2|\Delta|}$ for negative $\Delta$, and compare it with an exponential function. It shows that this alternate form of $U(\Delta)$ also decreases slower than exponentially, meaning that the ASH model still correctly predicts decreasing impatience.

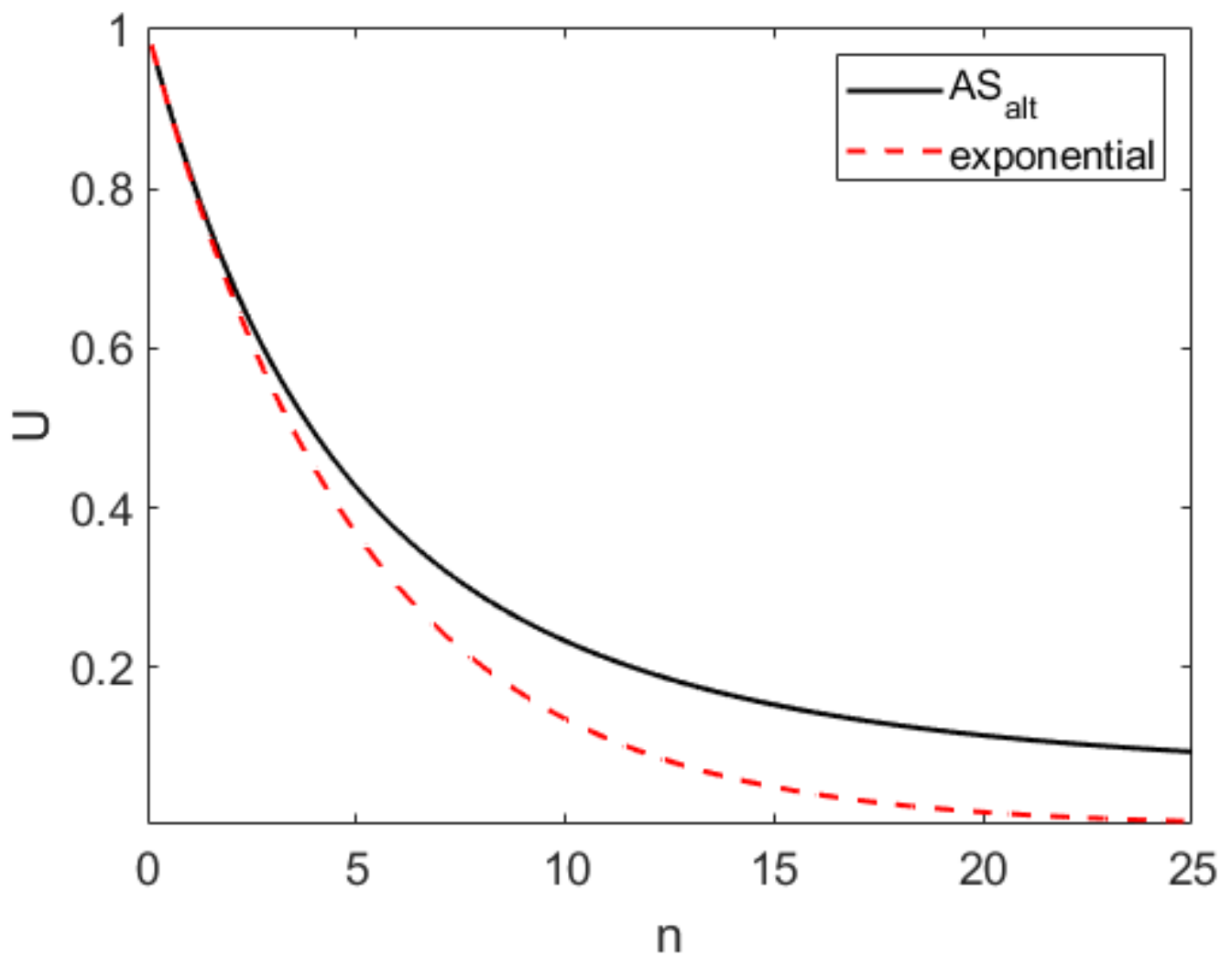

Figure A1: The discount factor $U$ for the alternative ASH model in which $U(\Delta) = e^{-k_2|\Delta|}$ for negative $\Delta$ and for exponential discounting ($U = e^{-k_{\text{exp}}n}$). $k_{\text{exp}} = 0.2$, $k_2 = 2$, $k = 3$, $\alpha = 1.6$, $p = 0.03$.

**Mathematical analysis for $|\Delta(n)|$**

*Small $n$*

From eq. (9), $|\Delta(n)| = Cq^{n+\alpha'}\frac{1-q^{n\alpha'}}{1-q^{\alpha'}}$, where $q = 1 - p \leq 1$. As in Chan & Toyoizumi (2024), we assume that the surprise function is weakly convex such that $0 < \alpha' \leq 1$.

To determine the model behavior for small $n$, we apply Taylor expansion around $n = 0$, where

$q^n = e^{n\ln q} = 1 + n\ln q + \frac{1}{2}n^2(\ln q)^2 + O(n^3)$, and

$1 - q^{n\alpha'} = 1 - e^{n\alpha'\ln q} = -\alpha'(\ln q)n - \frac{1}{2}\alpha'^2(\ln q)^2n^2 + O(n^3)$

Multiplying these two terms together and substituting them back into $|\Delta(n)|$ gives

$$|\Delta(n)| = -\left(\frac{C\alpha' q^{\alpha'} \ln q}{1-q^{\alpha'}}\right) n - \left(\frac{C\alpha'\left(1+\frac{\alpha'}{2}\right) q^{\alpha'} (\ln q)^2}{1-q^{\alpha'}}\right) n^2 + O(n^3), \tag{A1}$$

which suggests that $|\Delta(n)|$ is approximately linear at small $n$.

To show that the domain of near linearity of $|\Delta(n)|$ typically extends beyond the immediate proximity to $n = 0$ in realistic intertemporal decision-making scenarios, we note that the hazard probability $p$ is typically small because events that cause reward loss are unlikely within a single time step. Under this condition, we use the following approximations:

$$1 - q^{\alpha'} = 1 - (1-p)^{\alpha'} \approx \alpha' p,$$

$$1 - q^{n\alpha'} = 1 - e^{-n\alpha' p} \approx n\alpha' p \left(1 - \tfrac{1}{2} n\alpha' p\right)$$

$$q^{n+\alpha'} \approx q^{\alpha'} e^{-np} \approx e^{-np} \approx 1 - np$$

Substituting these approximations into eq. (9) gives

$$|\Delta(n)| \approx C(1-np)\frac{n\alpha' p\left(1-\frac{1}{2}n\alpha' p\right)}{\alpha' p} = Cn\left[1 - \left(1 + \frac{\alpha'}{2}\right) np + O[(np)^2]\right] \tag{A2}$$

Eq. (A2) suggests that $|\Delta(n)|$ remains approximately linear when $np$ is small. This also shows that the hyperbolic discounting exhibited in the ASH model is unaffected by the choice of time-step length, since reducing the step size results in a proportional decrease in the hazard probability per step, keeping the product $np$ invariant.

To further clarify that the utility $U(n)$ is approximately hyperbolic at small $np$ and begins to deviate at $n \sim \frac{1}{p}$, we rewrite eq. (9) as

$$|\Delta(n)| = Aq^n\left(1 - q^{\alpha' n}\right), \tag{A3}$$

where $A \equiv \frac{Cq^{\alpha'}}{1-q^{\alpha'}}$.

Substituting eq. (A3) into eq. (7) gives

$$[U(n)]^{-1} = \frac{1+k_2|\Delta(n)|}{q^n} = q^{-n} + k_2 A\left(1 - q^{\alpha' n}\right), \tag{A4}$$

where $q^n$ corresponds to the reduced expected reward due to hazard.

Under the assumption of $p \ll 1$, we have $\ln(q) \approx -p$ such that $q^{-n} \approx e^{np}$ and $q^{\alpha' n} \approx e^{\alpha' np}$. By applying Taylor expansions, we obtain $q^{-n} \approx 1 + np + O[(np)^2]$ and $1 - q^{\alpha' n} \approx \alpha' np + O[(np)^2]$. Substituting them into eq. (A4) gives

$$[U(n)]^{-1} \approx 1 + k' n + O[(np)^2], \tag{A5}$$

where $k' = (1 + k_2 A\alpha')p$.

It is clear from eq. (A5) that when $np$ is small, the higher-order terms $O[(np)^2]$ vanish, leading to $U(n) \approx \frac{1}{1+k'n}$ being approximately hyperbolic; whereas when $n \sim \frac{1}{p}$, $O[(np)^2]$ becomes non-negligible, and $U(n)$ transitions away from the hyperbolic form.

*Large $n$*

When $np$ is large, using the approximation $q = 1 - p \approx e^{-p}$ and $1 - q^{\alpha'} \approx \alpha' p$, eq. (9) becomes

$$|\Delta(n)| \approx \frac{C}{\alpha' p} e^{-u}\left(1 - e^{-\alpha' u}\right), \tag{A6}$$

where $u = np$.

The linear rise bends downward due to the competing exponential decay factor $e^{-u}$. Intuitively, as $n$ becomes large, the increase in negative surprise resulting from prolonged hazard exposure is offset by the reduced probability of surviving to later stages, where the largest negative surprises occur.

As $u = np \to \infty$, $|\Delta(n)| \to 0$. Note that the ASH model may not be realistic in this regime, as DMs cannot mentally simulate such an arbitrarily long sequence of events due to cognitive limits. Furthermore, $U$ becomes very small as a result of high cumulative hazard probability, rendering the discounted reward negligible and the exact functional form of discounting behaviorally inconsequential. Regardless of whether the non-monotonicity of $|\Delta(n)|$ in eq. (9) and eq. (A6) at large $n$ is biophysically realistic and behaviorally relevant, we show below that $\left|\frac{d|\Delta|}{dn}\right|$ remains small at large $n$ compared to $\left|\frac{d|\Delta|}{dn}\right|$ at small $n$, such that the ASH model predicts near-exponential discounting in this regime.

Differentiating eq. (A6) with respect to $n$ gives

$$\frac{d|\Delta|}{dn} = \frac{C}{\alpha'}\left[(1 + \alpha')e^{-(1+\alpha')u} - e^{-u}\right] \tag{A7}$$

To find the maximum value of $\left|\frac{d|\Delta|}{dn}\right|$ in the (decreasing) tail portion of $|\Delta(n)|$ as a function of $\alpha'$ ($M(\alpha')$), we compute the inflection point $u_i$ by solving $\frac{d^2|\Delta|}{dn^2} = 0$.

$$\frac{d^2|\Delta(u_i)|}{dn^2} = \frac{Cp}{\alpha'}\left[-(1 + \alpha')^2 e^{-(1+\alpha')u_i} + e^{-u_i}\right] = 0$$

$$\Rightarrow u_i = \frac{2\ln(1+\alpha')}{\alpha'} \tag{A8}$$

$M(\alpha')$ can be obtained by substituting $u = u_i$ into eq. (A7)

$$M(\alpha') = \left|\frac{C}{\alpha'}\left[(1 + \alpha')e^{-(1+\alpha')u_i} - e^{-u_i}\right]\right|$$

$$= \frac{C}{\alpha'} \left| (1+\alpha')^{-\left(\frac{2}{\alpha'}+1\right)} - (1+\alpha')^{-\frac{2}{\alpha'}} \right|$$

$$= C \frac{1}{(1+\alpha')^{1+\frac{2}{\alpha'}}} \tag{A9}$$

It can be easily shown that $M(\alpha')$ is decreasing with respect to $\alpha'$, and $M(\alpha' = 0^+) = Ce^{-2} \approx 0.135C$ (e.g. by L'Hôpital's rule). From eq. (A2), $\frac{d|\Delta|}{dn} = C$ at small $n$. Thus, the value of $\left|\frac{d|\Delta|}{dn}\right|$ is insubstantial in the large $n$ regime.

# Appendix A2: Analysis on timing risk

**Timing risk-seeking for expected-utility based hazard models**

First, we will show that the expected value of a time lottery with 2 possible reward timings, at $t = n - 1$ with probability $p_{\text{tr}}$ and at $t = n + 1$ with probability $1 - p_{\text{tr}}$, denoted by $E_{\text{tr}}$, is larger than that of a reward with a fixed delay at the weighted average time of possible reward timing in the abovementioned time lottery (i.e. $t = p_{\text{tr}}(n-1) + (1-p_{\text{tr}})(n+1)$), denoted by $E_{\text{fix}}$, in the context of the ASH model as depicted in Figure 4 of the main text.

$$E_{\text{tr}} - E_{\text{fix}} = q^{n-1}[p_{\text{tr}} + (1-p_{\text{tr}})q^2 - q^{2-2p_{\text{tr}}}], \tag{A10}$$

where $q = 1 - p$ as a reminder.

Let $H(p_{\text{tr}}) = p_{\text{tr}} + (1-p_{\text{tr}})q^2 - q^{2-2p_{\text{tr}}}$, it is trivial that $H(0) = H(1) = 0$. Computing the first and second derivatives of $H$ gives

$$H'(p_{\text{tr}}) = 1 - q^2 + 2\ln q \, e^{(1-p_{\text{tr}})\ln q^2} \tag{A11}$$

$$H''(p_{\text{tr}}) = -(2\ln q)^2 e^{2(1-p_{\text{tr}})\ln q} \, . \tag{A12}$$

Since $H''(p_{\text{tr}}) < 0$ and both $H$ and $H'(p_{\text{tr}})$ are continuous at $p_{\text{tr}} \in [0,1]$ and $p \in (0,1)$, it follows that $H$ and hence $E_{\text{tr}} - E_{\text{fix}} > 0$.

**The anticipated surprise for the ASH model**

We next derive the anticipated surprise for the timing-risk ($\Delta_{\text{tr}}$) and the fixed reward option ($\Delta_{\text{fix}}$), which are used to plot Figure 5.

One can easily obtain $\Delta_{\text{fix}}$ by making a change of variable $n \to p_{\text{tr}}(n-1) + (1-p_{\text{tr}})(n+1)$ in eq. (9). We are not showing the expression here.

For $\Delta_{\text{tr}}$, we first divide it into 3 parts as shown in Figure 4 in the main text: $\Delta_{\text{common}}$, $\Delta_{\text{tr0}}$, and $\Delta_{\text{late}}$, where $\Delta_{\text{common}} = \sum_{t=1}^{n-1} \Delta_t$ and $\Delta_{\text{late}} = \Delta_n + \Delta_{n+1}$. By the basic definition of anticipated surprise shown in eq. (1)-(4),

$$\Delta_{\text{tr0}} = (1-p)^{n-1}\{p_{\text{tr}}[1 - p_{\text{tr}} - (1-p_{\text{tr}})q^2]^\alpha - k(1-p_{\text{tr}})[p_{\text{tr}} - p_{\text{tr}}q^2]^\alpha\}$$

$$= q^{n-1}(1-q^2)^{\alpha}[p_{\text{tr}}(1-p_{\text{tr}})^{\alpha} - k(1-p_{\text{tr}})p_{\text{tr}}{}^{\alpha}], \tag{A13}$$

where the reference point is $p_{\text{tr}} + (1-p_{\text{tr}})\, q^2$.

Eq. (A13) can further be written as follows for easier interpretation:

$$\Delta_{\text{tr0}} = p_{\text{survival}} E_{\text{drop}}{}^{\alpha} \Delta_{\text{timing}}, \tag{A14}$$

where $p_{\text{survival}} = q^{n-1}$ corresponds to the probability of surviving hazard to the reward-timing resolution step, $E_{\text{drop}} = p_{\text{tr}}(1-q^2)$ corresponds to the drop in the expected reward when late reward delivery is confirmed, and $\Delta_{\text{timing}} = \frac{p_{\text{tr}}(1-p_{\text{tr}})^{\alpha} - k(1-p_{\text{tr}})p_{\text{tr}}{}^{\alpha}}{p_{\text{tr}}{}^{\alpha}}$ describes the surprise associated with the reward-timing distribution.

For $\Delta_{\text{common}}$, note that this is equivalent to the corresponding surprise in the simple intertemporal problem in Figure 2, except that the expected value at each step is scaled by a factor of $\frac{q^2(1-p_{\text{tr}})+p_{\text{tr}}}{q}$ and the number of time steps is shortened to $n-1$. The surprise can be computed by making the corresponding changes to eq. (8) and (9):

$$\Delta_{\text{common}} = \sum_{t=1}^{n-1} q^{t-1}\left(-Cq^{\alpha(n-t+1)}\right) \mathrm{E}_{\text{scale}}$$

$$= -C\mathrm{E}_{\text{scale}} q^{n+\alpha'}\left(\frac{1-q^{n\alpha'}}{1-q^{\alpha'}} - 1\right), \tag{A15}$$

where $\mathrm{E}_{\text{scale}} = \left(\frac{q^2(1-p_{\text{tr}})+p_{\text{tr}}}{q}\right)^{\alpha}$

For $\Delta_{\text{late}}$, instead of the reference expected value, it is the probability of entering the branch that has changed from the basic problem. In this particular case, since it only consists of two terms, it is easy to write them out directly using the definition of anticipated surprise:

$$\Delta_{\text{late}} = pq^{n+\alpha}(1-p_{\text{tr}})\left(p^{\alpha-1}(1+q^{1-\alpha}) - k(1+q^{\alpha-1})\right). \tag{A16}$$

The total surprise $\Delta_{\text{tr}}$ is the weighted sum of these components

$$\Delta_{\text{tr}} = \Delta_{\text{haz,tr}} + K_{\text{tr}}\Delta_{\text{tr0}}, \tag{A17}$$

where $\Delta_{\text{haz,tr}} = \Delta_{\text{common}} + \Delta_{\text{late}}$, and $K_{\text{tr}}$ is the weighting of the timing-risk driven surprise relative to hazard-driven surprise.

# Appendix A3: Analysis on dual-risk discounting

### Derivation of the discount factor $D_{\text{A}}$ and $D_{\text{B}}$

We will first derive the discount factors for scheme A ($D_{\text{A}}$) and Scheme B ($D_{\text{B}}$) shown in eq. (15) and (16). The surprise for the hazard risk-only option ($\Delta_{\text{haz}}$) is given by eq. (9), which we will restate here:

$$\Delta_{\text{haz}}(x) = -C(x)(1-x)^{n+\alpha'}\frac{1-(1-x)^{n\alpha'}}{1-(1-x)^{\alpha'}}, \tag{A18}$$

where $C(x) = kx - x^{\alpha}(1-x)^{\alpha}$.

The surprise for the probabilistic risk-only option ($\Delta_{\text{pr}}$) can be computed by eq. (1)-(4):

$$\Delta_{\text{pr}} = p_{\text{pr}}(1-p_{\text{pr}})^{\alpha} - k(1-p_{\text{pr}})p_{\text{pr}}{}^{\alpha} \tag{A19}$$

For the dual risk option resolved using Scheme A, assuming the probabilistic risk is resolved after the hazard risk, ($\Delta_{\text{pta}}$), we divide the surprise into 2 components: $\Delta_{\text{haz,pt}}$ for the surprise induced during the resolution of hazard risk and $\Delta_{\text{pr,pt}}$ for probabilistic risk, and $\Delta_{\text{pta}} = \Delta_{\text{haz,pt}} + \Delta_{\text{pr,pt}}$

Note that under Scheme A, the resolution of hazard risk in a dual-risk option is equivalent to the resolution in a hazard risk-only option, except that the expected value at each step is scaled by a factor of $\text{p}_{\text{pr}}$, meaning

$$\Delta_{\text{haz,pt}} = p_{\text{pr}}{}^{\alpha}\Delta_{\text{haz}} \tag{A20}$$

By making similar comparisons for the probabilistic risk component, we obtain

$$\Delta_{\text{pr,pt}} = q^{n}\Delta_{\text{pr}} \tag{A21}$$

As a reminder, $q = 1 - p$. The generic discount factor in eq. (10) can be further simplified using eq. (6)

$$D = \frac{U_{\text{pt}}}{U_{\text{p}}U_{\text{t}}} = \frac{U_{0,\text{pt}}g(\Delta_{\text{pt}})}{U_{0,\text{p}}U_{0,\text{t}}g(\Delta_{\text{p}})g(\Delta_{\text{t}})} = \frac{g(\Delta_{\text{pt}})}{g(\Delta_{\text{p}})g(\Delta_{\text{t}})}, \tag{A22}$$

where $U_{0,\text{pt}}$, $U_{0,\text{p}}$, $U_{0,\text{t}}$ refers to the (reduced) expected value for the dual-risk, probabilistic risk-only and hazard risk-only options, which all got cancelled out.

$D_{\text{A}}$ can then be obtained by substituting $\Delta_{\text{pta}}$ into eq. (A22) and applying the appropriate scaling:

$$D_{\text{A}} = \frac{g\left(K_{\text{pr}}q^{n}\Delta_{\text{pr}} + K_{\text{haz}}p_{\text{pr}}{}^{\alpha}\Delta_{\text{haz}}(p)\right)}{g\left(K_{\text{pr}}\Delta_{\text{pr}}\right)g\left(K_{\text{haz}}\Delta_{\text{haz}}(p)\right)} \tag{A23}$$

In the case of probabilistic risk resolving before the hazard risk, the anticipated surprise ($\Delta'_{\text{pta}}$) can be easily obtained by appropriately scaling $\Delta_{\text{haz}}$ and $\Delta_{\text{pr}}$ to reflect the prior probability of late branches and the update of expected value in intermediate states:

$$\Delta'_{\text{pta}} = p_{\text{pr}}\Delta_{\text{haz}} + q^{n\alpha}\Delta_{\text{pr}}. \tag{A24}$$

The resulting discount factor can likewise be obtained substituting $\Delta_{\text{pta}}$ into eq. (A22) (not shown here).

For the dual risk option in Scheme B, the surprise $\Delta_{\mathrm{ptb}}$ can be obtained by making a change of variable $p \to p'$ in eq. (9), where $p' = 1 - \sqrt[n]{p_{\mathrm{pr}}}(1-p)$, and substitute it into $\Delta_{\mathrm{haz}}$, i.e. $\Delta_{\mathrm{ptb}} = \Delta_{\mathrm{haz}}(p')$. Since the risk has become 'hazard' in nature, it is scaled by $K_{\mathrm{haz}}$, leading to:

$$D_{\mathrm{B}} = \frac{g(K_{\mathrm{haz}}\Delta_{\mathrm{haz}}(p'))}{g(K_{\mathrm{pr}}\Delta_{\mathrm{pr}})g(K_{\mathrm{haz}}\Delta_{\mathrm{haz}}(p))} \tag{A25}$$

**Analysis of $D_{\mathrm{A}}$ under generic form of $g$ and the specific form in eq. (7)**

Here, we show that a sufficient condition for $D_{\mathrm{A}} > 1$ is that $-\ln(g(\Delta))$ is concave. $-\ln(g(\Delta))$ being concave implies that $f(\Delta) = \ln(g(\Delta))$ is convex, which, by Jensen's inequality, means

$$f(\lambda x_1 + (1-\lambda)x_2) \leq \lambda f(x_1) + (1-\lambda)f(x_2) \; \forall \lambda \epsilon [0,1], \tag{A26}$$

Substituting $\lambda = \frac{y_1}{y_1+y_2}$, $x_1 = y_1 + y_2$, $x_2 = 0$, where $y_1, y_2 < 0$, into eq. (A26) yields

$$f(y_1) \leq \frac{y_1}{y_1+y_2} f(y_1+y_2) + \left(1 - \frac{y_1}{y_1+y_2}\right) f(0) = \frac{y_1}{y_1+y_2} f(y_1+y_2), \tag{A27}$$

Note that $f(0) = \ln\big(g(0)\big) = 0$, since $g(0) = 1$ (see Section 2). Similarly, by substituting $\lambda = \frac{y_2}{y_1+y_2}$ into eq. (A26) instead, we have

$$f(y_2) \leq \frac{y_2}{y_1+y_2} f(y_1+y_2). \tag{A28}$$

Adding eq. (A27) and (A28) gives

$$f(y_1+y_2) \geq f(y_1) + f(y_2). \tag{A29}$$

Substituting $y_1 = K_{\mathrm{pr}}\Delta_{\mathrm{pr}}$ and $y_2 = K_{\mathrm{haz}}\Delta_{\mathrm{haz}}(p)$ into eq. (A29) gives

$$\ln\left[g\left(K_{\mathrm{pr}}\Delta_{\mathrm{pr}} + K_{\mathrm{haz}}\Delta_{\mathrm{haz}}(p)\right)\right] \geq \ln\left[g\left(K_{\mathrm{pr}}\Delta_{\mathrm{pr}}\right)\right] + \ln\left[g\left(K_{\mathrm{haz}}\Delta_{\mathrm{haz}}(p)\right)\right] \tag{A30}$$

Rearranging terms gives $\ln\left[\frac{g(K_{\mathrm{pr}}\Delta_{\mathrm{pr}}+K_{\mathrm{haz}}\Delta_{\mathrm{haz}}(p))}{g(K_{\mathrm{pr}}\Delta_{\mathrm{pr}})g(K_{\mathrm{haz}}\Delta_{\mathrm{haz}}(p))}\right] \geq 0$, or equivalently

$$\frac{g(K_{\mathrm{pr}}\Delta_{\mathrm{pr}}+K_{\mathrm{haz}}\Delta_{\mathrm{haz}}(p))}{g(K_{\mathrm{pr}}\Delta_{\mathrm{pr}})g(K_{\mathrm{haz}}\Delta_{\mathrm{haz}}(p))} \geq 1 \tag{A31}$$

Since ${p_{\mathrm{pr}}}^{\alpha}, q^n < 1$, $\Delta_{\mathrm{haz}}(p), \Delta_{\mathrm{pr}} < 0$ and $g(\Delta)$ is decreasing with $|\Delta|$ for $\Delta < 0$ (See eq. (6)), we have $g\big(K_{\mathrm{pr}}\Delta_{\mathrm{pr}} + K_{\mathrm{haz}}\Delta_{\mathrm{haz}}(p)\big) < g\big(K_{\mathrm{pr}}q^n\Delta_{\mathrm{pr}} + K_{\mathrm{haz}}{p_{\mathrm{pr}}}^{\alpha}\Delta_{\mathrm{haz}}(p)\big)$. Substituting this into eq. (A31) and comparing with eq. (A23), we have

$$1 \leq \frac{g\big(K_{\mathrm{pr}}\Delta_{\mathrm{pr}}+K_{\mathrm{haz}}\Delta_{\mathrm{haz}}(p)\big)}{g(K_{\mathrm{pr}}\Delta_{\mathrm{pr}})g(K_{\mathrm{haz}}\Delta_{\mathrm{haz}}(p))} < \frac{g\big(K_{\mathrm{pr}}q^n\Delta_{\mathrm{pr}}+K_{\mathrm{haz}}{p_{\mathrm{pr}}}^{\alpha}\Delta_{\mathrm{haz}}(p)\big)}{g(K_{\mathrm{pr}}\Delta_{\mathrm{pr}})g(K_{\mathrm{haz}}\Delta_{\mathrm{haz}}(p))} = D_{\mathrm{A}}. \tag{A32}$$

In eq. (7), we propose the specific form $g(\Delta) = \frac{1}{1+k_2|\Delta|}$ for $\Delta < 0$. Substituting it into eq. (A23) gives

$$D_{\mathrm{A}} = \frac{(1+k_{\mathrm{eff}}|\Delta_{\mathrm{pr}}|)(1+Rk_{\mathrm{eff}}|\Delta_{\mathrm{haz}}(p)|)}{1+k_{\mathrm{eff}}(q^n|\Delta_{\mathrm{pr}}|+Rp_{\mathrm{pr}}{}^{\alpha}|\Delta_{\mathrm{haz}}(p)|)}, \quad \text{(A33)}$$

where $k_{\mathrm{eff}} = k_2 K_{\mathrm{pr}}$ and $R = \frac{K_{\mathrm{haz}}}{K_{\mathrm{pr}}}$.

It is straightforward to show directly from (A33) that $D_{\mathrm{A}} > 1$:

$$D_{\mathrm{A}} > \frac{(1+k_{\mathrm{eff}}|\Delta_{\mathrm{pr}}|)(1+Rk_{\mathrm{eff}}|\Delta_{\mathrm{haz}}(p)|)}{1+k_{\mathrm{eff}}|\Delta_{\mathrm{pr}}|+Rk_{\mathrm{eff}}|\Delta_{\mathrm{haz}}(p)|} = 1 + \frac{Rk_{\mathrm{eff}}{}^2|\Delta_{\mathrm{pr}}||\Delta_{\mathrm{haz}}(p)|}{1+k_{\mathrm{eff}}|\Delta_{\mathrm{pr}}|+Rk_{\mathrm{eff}}|\Delta_{\mathrm{haz}}(p)|} > 1 \quad \text{(A34)}$$

Alternatively, we can easily show that $\ln(g(\Delta))$ is convex for this $g$ by taking the second derivatives with respect to $\Delta$:

$$\frac{d}{d\Delta}\ln(g(\Delta)) = \frac{k_2}{1+k_2|\Delta|} \text{ and } \frac{d^2}{d\Delta^2}\ln(g(\Delta)) = \frac{k_2{}^2}{(1+k_2|\Delta|)^2} > 0. \quad \text{(A35)}$$

**Analysis of $D_B$ under the specific form of $g$ in eq. (7)**

We first show that for $g(\Delta) = \frac{1}{1+k_2|\Delta|}$, $\frac{d}{dK_{\mathrm{haz}}}\ln(D_{\mathrm{B}}) < 0$, where $D_{\mathrm{B}}$ is given by eq. (16) and eq. (A25). This is equivalent to showing that $|\Delta_{\mathrm{haz}}(\mathrm{x})|h(\Delta_{\mathrm{haz}}(x))$ decreases with $x$, where $h(z) = \frac{1}{g(z)}\frac{dg(\mathrm{z})}{d|\mathrm{z}|}$ (See Section 4.3 in the main text). Substituting this form of $g$ into $|\Delta_{\mathrm{haz}}(\mathrm{x})|h(\Delta_{\mathrm{haz}}(x))$ yields

$$|\Delta_{\mathrm{haz}}(\mathrm{x})|h\big(\Delta_{\mathrm{haz}}(x)\big) = |\Delta_{\mathrm{haz}}(\mathrm{x})|[1+k_2|\Delta_{\mathrm{haz}}(\mathrm{x})|]\frac{d}{d|\Delta_{\mathrm{haz}}(\mathrm{x})|}\left(\frac{1}{1+k_2|\Delta_{\mathrm{haz}}(\mathrm{x})|}\right) = \frac{-k_2|\Delta_{\mathrm{haz}}(\mathrm{x})|}{1+k_2|\Delta_{\mathrm{haz}}(\mathrm{x})|} \quad \text{(A36)}$$

Taking the derivative of eq. (A36) with respect to x yields

$$\frac{d}{dx}\frac{-k_2|\Delta_{\mathrm{haz}}(\mathrm{x})|}{1+k_2|\Delta_{\mathrm{haz}}(\mathrm{x})|} = \frac{d}{dx}\Big(1-\frac{1}{1+k_2|\Delta_{\mathrm{haz}}(\mathrm{x})|}\Big) = \frac{-k_2}{(1+k_2|\Delta_{\mathrm{haz}}(\mathrm{x})|)^2}\frac{d|\Delta_{\mathrm{haz}}(\mathrm{x})|}{dx} < 0. \quad \text{(A37)}$$

The last inequality in eq. (A37) holds because $|\Delta_{\mathrm{haz}}(\mathrm{x})|$ increases with x (See Figure 3 and Appendix A1) in the regime of our interest (non-small $p_{\mathrm{pr}}$ and small $np$), i.e. $\frac{d|\Delta_{\mathrm{haz}}(\mathrm{x})|}{dx} > 0$. Eq. (A37) means that $|\Delta_{\mathrm{haz}}(\mathrm{x})|h(\Delta_{\mathrm{haz}}(x))$ decreases with $x$, and thus $\frac{d}{dK_{\mathrm{haz}}}\ln(D_{\mathrm{B}}) < 0$.

We next show the condition for $D_{\mathrm{B}} < 1$. Substituting $g(\Delta) = \frac{1}{1+k_2|\Delta|}$ into eq. (A25) gives

$$D_{\mathrm{B}} = \frac{(1+k_{\mathrm{eff}}|\Delta_{\mathrm{pr}}|)(1+Rk_{\mathrm{eff}}|\Delta_{\mathrm{haz}}(p)|)}{1+Rk_{\mathrm{eff}}|\Delta_{\mathrm{haz}}(p')|} \quad \text{(A38)}$$

$D_{\mathrm{B}} < 1 \Leftrightarrow 1 + Rk_{\mathrm{eff}}|\Delta_{\mathrm{haz}}(p')| > (1+k_{\mathrm{eff}}|\Delta_{\mathrm{pr}}|)(1+Rk_{\mathrm{eff}}|\Delta_{\mathrm{haz}}(p)|)$. By rearranging the terms, we obtain the following: $D_{\mathrm{B}} < 1$ if and only if

$$|\Delta_{\mathrm{haz}}(p')| > |\Delta_{\mathrm{haz}}(p)|\big(1+k_{\mathrm{eff}}|\Delta_{\mathrm{pr}}|\big) + \frac{|\Delta_{\mathrm{pr}}|}{R}. \quad \text{(A39)}$$

# Appendix A4: Analysis on outcome scaling

Here, we demonstrate how to perform the outcome scaling described in Section 5.2.1 in the main text and the effects of the scaling.

Using the option (-1, 0.5; 1, 0.5) as an example, we first make the transformation on the outcome $x' = \frac{x+1}{2}$ to obtain an alternate option (0, 0.5; 1, 0.5). Then we compute the utility for this alternate option $U'$ (Using the parameters in Figure 1 gives $U' = 0.301$). Finally, we obtain the utility for the original problem $U$ by making the reverse transformation $U = 2U' - 1$ (For the parameters in Figure 1, we have $U = -0.398$, which is quite plausible empirically).

To further illustrate the effects of the treatments outlined in the last paragraph, let's consider the option (0, 0.5; $\frac{1}{p}$, $p$). In this problem, $E = 1$ independent of $p$, and the positive outcome could become very large to compensate for small $p$. Figure A3 plots the utility function versus $p$ with and without outcome transformation. Without transformation (Figure A3, left), there is a huge spike in $U$ for small $p$. Since the small $p$ regime corresponds to activities like buying lottery tickets, the spike implies that people would unrealistically spend an extremely large fortune on such activities. In addition, $U$ is also disproportionately suppressed at medium $p$. By contrast, there is no spike at the small $p$ regime after performing the transformation (Figure A3, right), and $U$ stays reasonably close to the expected value 1.

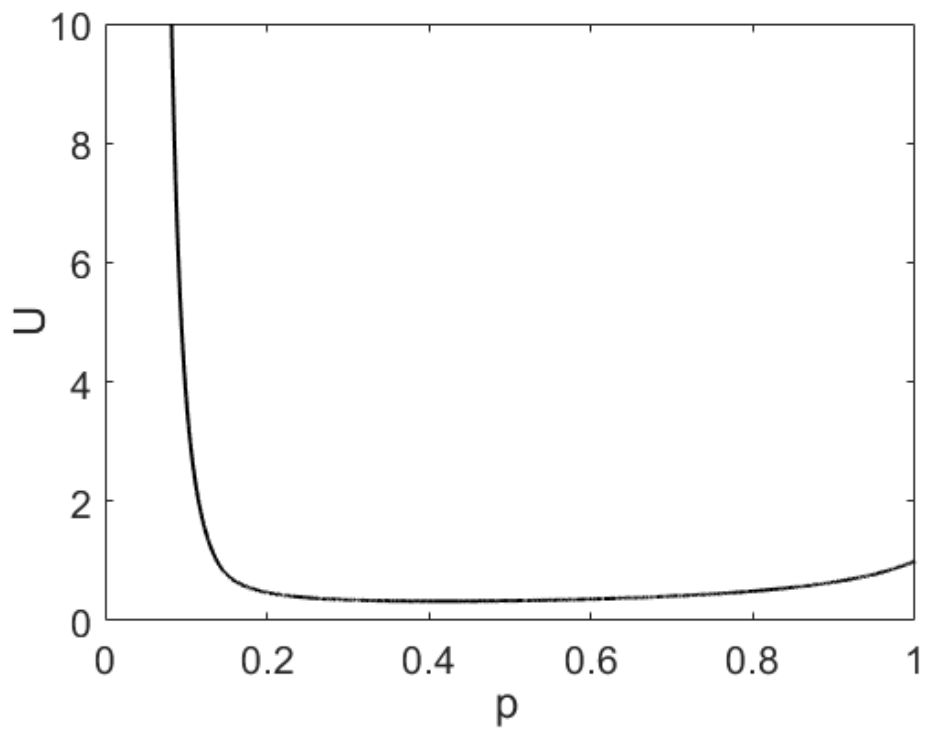

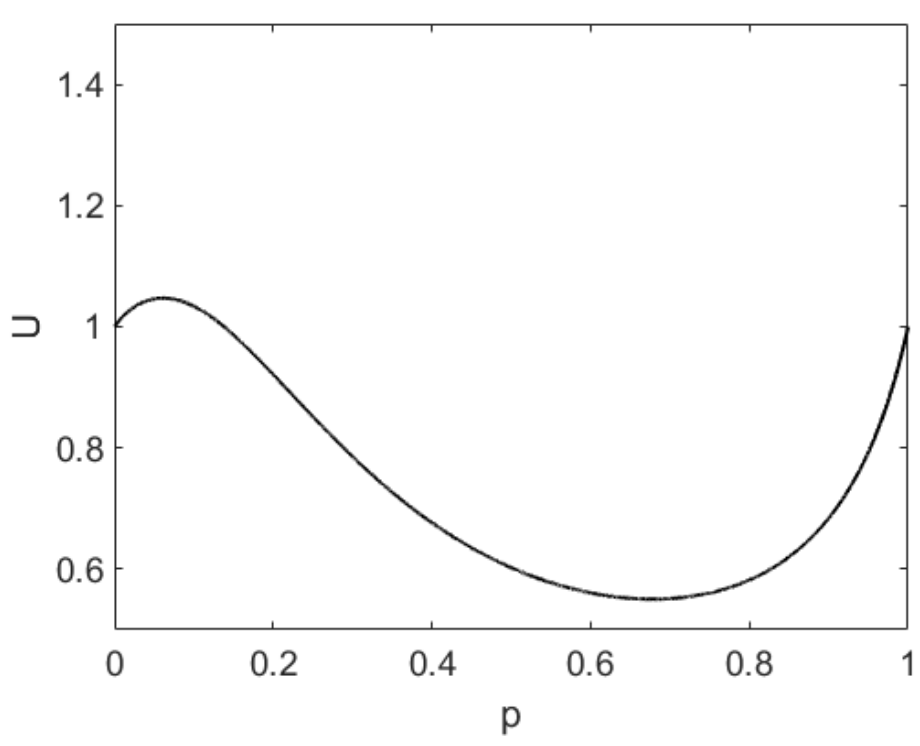

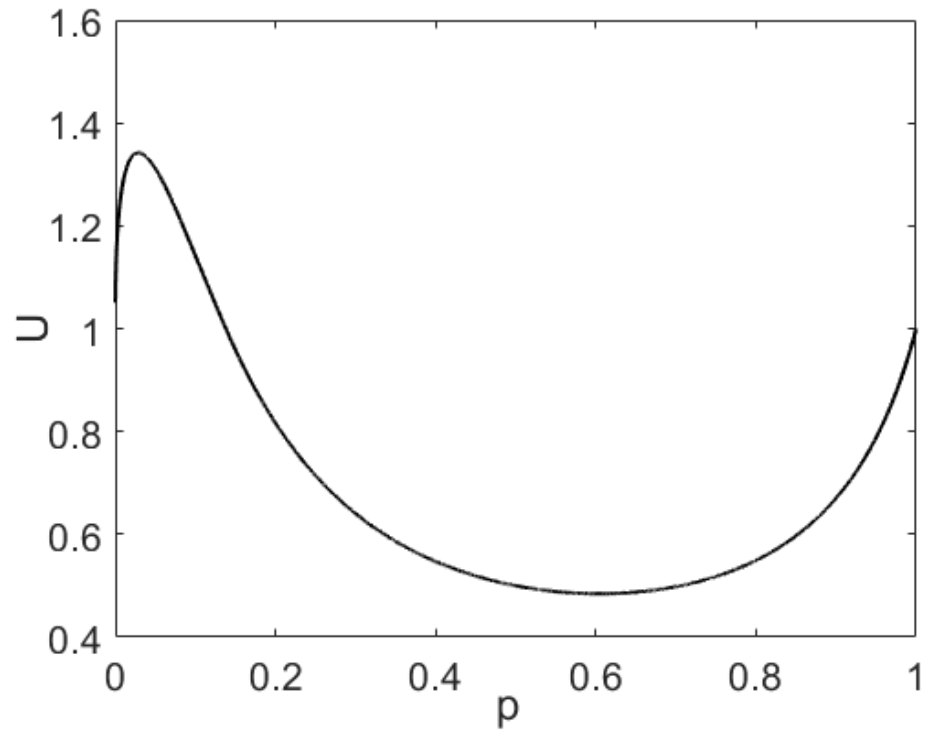

Figure A3: The utility for the problem $(0, 0.5; \frac{1}{p}, p)$ when the outcome is not scaled (top left), fully scaled (top right) and incompletely scaled (bottom left). The parameters are the same as the ones used in Figure 1.

Note that the risk seeking behavior at small $p$ is unrealistically strongly attenuated by the transformation, as it completely removes the magnitude effects empirically observed. For the specific option $(0, 0.5; \frac{1}{p}, p)$, a realistic utility would most likely lie somewhere between top left and top right of Figure A3. One way to reproduce such a 'realistic utility' is to perform a 'partial scaling'. In Figure A3 (bottom left), we use the scaling $x' = p^{\frac{1}{\alpha}}x$, transforming the problem into $(0, 0.5; p^{1-\frac{1}{\alpha}}, p)$. With this partial scaling, a 30-40% increase of utility is generated at small $p$, this is likely to be quite realistic given that the payouts of real-life lotteries are often undercut by a similar amount (due to various reasons like taxes and operating costs).

More detailed and rigorous analysis of the outcome scaling in the ASH model shall be left for future studies.